\documentclass[manuscript]{acmart}

\usepackage{CJKutf8}
\usepackage[utf8]{inputenc}
\usepackage[T1]{fontenc}
\usepackage{textcomp}
\usepackage{xcolor,colortbl}
\usepackage{multirow}
\usepackage{multicol}
\usepackage{threeparttable}
\usepackage{soul}
\usepackage{hyperref} 
\usepackage{tabularx}
\usepackage{longtable}
\usepackage{subfig}
\usepackage{overpic}
\usepackage{array}
\usepackage{makecell}
\usepackage[most]{tcolorbox}
\tcbset{
  mytakeawaybox/.style={
    colback=gray!10,    
    colframe=black,     
    boxrule=0.5pt,      
    arc=2mm,            
    left=5pt, right=5pt, top=5pt, bottom=5pt,
    fonttitle=\bfseries,
    title=Key Findings
  }
}

\AtBeginDocument{%
  \providecommand\BibTeX{{%
    \normalfont B\kern-0.5em{\scshape i\kern-0.25em b}\kern-0.8em\TeX}}}

\newcommand{\modtext}[1]{\textcolor{black}{#1}}
\begin{document}
\setcopyright{acmcopyright}
\copyrightyear{2021}
\acmYear{2021}
\acmDOI{XX.XXXX/XXXXXXX.XXXXXXX}

\acmJournal{JACM}
\acmVolume{1}
\acmNumber{1}
\acmArticle{1}
\acmMonth{1}

% \title{Factors of User Acceptance and Trust towards Distributed Protocols for Fairness Monitoring}

%\title{Navigating User Acceptance of Multi-party Computation Protocol Design of Data sharing for Fairness Monitoring}
\title[Examining User Acceptance of Multi-Party Computation Protocols for Fairness Monitoring]{When Feasibility of Fairness Audits Relies on Willingness to Share Data: Examining User Acceptance of Multi-Party Computation Protocols for Fairness Monitoring}

\author{Changyang He}
\authornote{These authors contributed equally to this work.}
\authornote{This work was primarily conducted during the author's stay at the Max Planck Institute for Security and Privacy (MPI-SP).}
\affiliation{%
  \institution{Harbin Institute of Technology, Shenzhen}
  \city{Shenzhen}
  \country{China}
}
\affiliation{%
  \institution{Max Planck Institute for Security and Privacy (MPI-SP)}
  \city{Bochum}
  \country{Germany}
}
\email{hechangy@hit.edu.cn}

\author{Parnian Jahangirirad}
\authornotemark[1]
\affiliation{%
  \institution{Max Planck Institute for Security and Privacy (MPI-SP)}
  \city{Bochum}
  \country{Germany}
}
\affiliation{%
  \institution{Saarland University}
  \city{Saarbrücken}
  \country{Germany}
}
\email{parnian.jahangirirad@mpi-sp.org}

\author{Lin Kyi}
\affiliation{%
  \institution{Max Planck Institute for Security and Privacy (MPI-SP)}
  \city{Bochum}
  \country{Germany}
}
\email{lin.kyi@mpi-sp.org}

\author{Asia J. Biega}
\authornote{Corresponding author.}
\affiliation{%
  \institution{Max Planck Institute for Security and Privacy (MPI-SP)}
  \city{Bochum}
  \country{Germany}
}
\email{asia.biega@mpi-sp.org}

\renewcommand{\shortauthors}{He et al.}

\begin{abstract}

Fairness monitoring is critical for detecting algorithmic bias, as mandated by the EU AI Act. Since such monitoring requires sensitive user data (e.g., ethnicity), the AI Act permits its processing only with strict privacy measures, such as multi-party computation (MPC), in compliance with the GDPR. However, the effectiveness of such secure monitoring protocols ultimately depends on people's willingness to share their data. Little is known about how different MPC protocol designs shape user acceptance. To address this, we conducted an online survey with 833 participants in Europe, examining user acceptance of various MPC protocol designs for fairness monitoring. Findings suggest that users prioritized risk-related attributes (e.g., privacy protection mechanism) in direct evaluation but benefit-related attributes (e.g., fairness objective) in simulated choices, with acceptance shaped by their fairness and privacy orientations. We derive implications for deploying and communicating privacy-preserving protocols in ways that foster informed consent and align with user expectations.

\end{abstract}




\maketitle

\section{Introduction}

Fairness monitoring has become essential to ensure algorithmic accountability, particularly as technology regulations mandate non-discriminatory outcomes in high-stakes applications. The European Union's AI Act, for instance, designates employment-related systems as high-risk and requires providers to examine them for bias and discrimination~\cite{edwards2021eu}. However, this mandate creates a fundamental challenge: effective fairness measurement relies on access to sensitive user data such as ethnicity or sexual orientation~\cite{fabris2024fairness,raghavan2020mitigating,mehrabi2021survey}, yet this information is heavily protected under privacy regulations like the European Union's General Data Protection Regulation (GDPR)~\cite{voigt2017eu}. The GDPR designates such information as special categories of data and prohibits their processing by default, in particular without valid consent~\cite{hoofnagle2019european}. In hiring, as in many high-risk applications, it has been argued that obtaining valid (freely given) consent might be impossible due to the vulnerability and the inherent power differentials between job seekers and employers~\cite{breen2020gdpr,parviainen2022can,hunkenschroer2023ai}. 

To address this tension, the AI Act permits processing of sensitive data for fairness auditing under certain conditions (Art. 10(5)), including the deployment of ``state-of-the-art security and privacy-preserving measures''~\cite{van2024using}. 
Privacy-preserving solutions, such as pseudonymization, anonymization, differential privacy, or multi-party computation (MPC), have been proposed as technically viable bases for compliant monitoring approaches, enabling fairness metric computation without disclosing sensitive data~\cite{goldreich1998secure,helminger2022multi}. Out of these solutions, MPC in particular is a promising direction as it preserves data fidelity without noise injection -- which is particularly valuable in fairness monitoring scenarios where measurement has to cope with limited sample sizes~\cite{zaccour2025access}. 

% GDPR~\cite{voigt2017eu}.

However, the feasibility of fairness measurement ultimately depends on \modtext{end} users' willingness to share their sensitive data with service providers or auditors. Even if privacy-preserving methods can technically be applied to fairness monitoring, the effectiveness of such secure monitoring frameworks hinges on the acceptance and participation of end users~\cite{kacsmar2023comprehension,nanayakkara2023chances}, \modtext{who are typically non-experts in privacy-preserving computations}. Yet the challenge of selecting and communicating specific deployment designs for these protocols in ways that align with users' expectations remains largely unexplored. Recent HCI research has emphasized that privacy-preserving solutions often fall short without user-centered approaches that support trust, transparency, and meaningful consent~\cite{franzen2024communicating,dibia2024sok,xiong2020towards,xiong2022using,kuhtreiber2022replication,cummings2021need}. 
This user-centered perspective is particularly crucial in fairness monitoring, where building trust across diverse user groups is essential to ensure representative datasets for meaningful results~\cite{fabris2022algorithmic}.

To address this gap, we examine user acceptance of various MPC protocol designs for fairness monitoring. Applying Privacy Calculus Theory~\cite{culnan1999information}, we categorized design attributes into benefit-related attributes (\textit{fairness objective} and \textit{monetary incentive}), risk-related attributes (\textit{privacy protection mechanism} and \textit{data storage}), and attributes related to both perceived benefits and risks (\textit{collected information type}, \textit{monitoring actor} and \textit{data use}). We first investigate how users navigate the privacy-benefit trade-off when deciding whether to participate in fairness monitoring. Then, we examine how users' fairness orientations, privacy orientations,  as well as personal contexts and demographics, as key elements in privacy calculus \cite{seberger2021us,ghaiumy2024personalizing,woodruff2014would}, correlate with protocol acceptance. 
% Our results can inform for tailoring communication that supports meaningfully informed consent. In particular, we propose the following research questions:
We address the following research questions:
\begin{itemize}
    \item \textbf{RQ1}: What are users' priorities and preferences regarding different designs of MPC protocols for fairness monitoring?
    \item \textbf{RQ2}: How do users' fairness and privacy orientations and contexts correlate with user acceptance of MPC protocols for fairness monitoring?
\end{itemize}

To answer these research questions, we conducted an online survey with 833 \modtext{current job seekers} in European countries within the jurisdiction of the GDPR and the AI Act. \modtext{Job seekers represent potential data sharers in the context of fairness monitoring in algorithmic hiring, a representative data-sharing setting that requires sensitive data to detect algorithmic bias}. 

For RQ1, we measured \emph{stated} attribute importance for users through direct attribute ranking and \emph{revealed} attribute importance through scenario-based conjoint analysis across different design attributes. Our findings reveal that users prioritized attributes relevant to privacy risks such as \textit{collected information type} and \textit{privacy protection mechanisms} in direct attribute ranking, but focused on benefit-related attributes like \textit{fairness objective} and \textit{monetary incentive} in the scenario-based study. Users demonstrated several expected privacy preferences,  such as favoring additional distributed data storage over anonymization alone, or trusting research centers more than commercial companies. However, more extensive data requests and expanded data use surprisingly \emph{increased} user willingness to share data. We hypothesize that this is due to the perceived contributions of the data to the system's fairness goals. 

% These benefit-privacy trade-offs varied significantly across demographic groups, however.

For RQ2, through regression analysis, we found that users' fairness and privacy orientations shaped their protocol preferences: users with higher prior openness to data sharing emphasized fairness objectives, while privacy-oriented users (those employing active privacy safeguards and high valuation of data protection transparency) focused less on fairness objectives or monetary incentives and more on risk-related attributes. Based on these findings, we discuss implications for designing and communicating privacy-preserving fairness monitoring protocols under the GDPR and the AI Act.

% In summary, this work makes the following contributions to the HCI community: (1) we took a human-centered approach to investigate user acceptance of privacy-preserving protocol design for fairness monitoring, revealing the nuances of benefit-privacy trade-offs among different demographic groups; (2) we examined the role of users' fairness and privacy orientations and contexts in influencing user preference of protocol design, shedding light on how users' fairness and privacy orientations and contexts shape their privacy calculus for protocol design considerations; (3) we proposed practical implications for privacy-preserving fairness monitoring protocols, covering both inclusive protocol design to meet users' expectations and effective protocol communication for meaningfully informed consent.

% % Option to make it shorter
This work makes the following contributions to HCI: (1) a human-centered investigation of user acceptance of MPC fairness monitoring protocols that reveals nuanced benefit-privacy trade-offs across design attributes; (2) analysis of how users' orientations and contexts shape privacy calculus for protocol design; and (3) practical implications for deploying privacy-preserving protocols and communicating them to ensure meaningfully informed consent.

\section{Related Work}

\subsection{Privacy-Preserving Fairness Monitoring}

With the increasing adoption of algorithmic decision-making in high-stakes domains such as hiring and criminal justice, there have been significant concerns about the potential of data-driven algorithms to reinforce and amplify historical and social biases~\cite{kordzadeh2022algorithmic,mehrabi2021survey}. In response, researchers and practitioners have devoted substantial effort to improving algorithmic fairness, developing techniques such as bias mitigation~\cite{hort2024bias,ferrara2023fairness} and fairness-aware learning~\cite{le2022survey,fabris2024fairness}. Despite the progress of algorithmic fairness, algorithms in real-world deployments frequently face data and operational conditions that diverge substantially from the training environment (e.g., data drift~\cite{deho2024past} and concept drift~\cite{lu2018learning}), limiting the robustness and interpretability of fair algorithms. As such, recent studies have highlighted the importance of fairness monitoring as a continuous process of tracking algorithmic fairness in the post-deployment stage~\cite{fabris2024fairness,bhargavachallenges,bellamy2019ai}. This is consistent with the European Union's AI Act~\cite{EU_AI_Act_Article_72}, which mandates ``Post-Market Monitoring'' to ensure that high-risk AI systems remain compliant with regulatory requirements throughout their entire lifecycle.

% Although researchers and practitioners have employed diverse approaches throughout the algorithmic lifecycle, fairness monitoring.

Different from the training environment that largely assumes access to sensitive attributes given the publicly available datasets, fairness monitoring in the post-deployment stage inevitably faces privacy constraints in real-world settings~\cite{holstein2019improving,zaccour2025access,veale2017fairer,chang2021privacy,stadler2022search}. \modtext{For example, concerns about sensitive data pose significant challenges in New York City's audits for algorithmic bias~\cite{groves2024auditing}}. Therefore, privacy-preserving fairness monitoring becomes a crucial topic. For example, in the EU, it is strictly necessary for the purpose of ensuring bias detection and correction in relation to the high-risk AI systems, the providers of such systems may exceptionally process special categories of personal data under Article 10(5) of AI Act~\cite{van2024using}. However, such processing must be ``subject to appropriate safeguards for the fundamental rights and freedoms of natural persons''. Therefore, researchers in algorithmic fairness and privacy have proposed various privacy-preserving fairness monitoring protocols~\cite{zaccour2025access,chang2021privacy,islam2023differential}. Several notable examples are outlined below.

\paragraph{Differential Privacy (DP)} DP provides rigorous, mathematically formal guarantees for limiting the disclosure risk of individual data~\cite{chang2021privacy}. In the context of fairness monitoring, it can be applied by injecting carefully calibrated noise into aggregate statistics (e.g., confusion matrices or group-specific fairness metrics). This approach enables privacy-preserving assessments of fairness while preventing the leakage of raw data or exact model outputs~\cite{zaccour2025access,chang2021privacy,jagielski2019differentially}. Nonetheless, the addition of noise inevitably reduces data granularity and may compromise the reliability of fairness evaluations. Zaccour et~al.\ found that DP-protected aggregates preserve fairness-metric accuracy only when samples (n) are large and privacy budgets ($\varepsilon$) aren’t extreme, e.g., reliable at $\varepsilon \approx 0.5$ with $n = 1{,}000$ or at $\varepsilon \approx 0.05$ with $n > 5{,}000$; smaller $n$ or very low $\varepsilon$ makes estimates unreliable
~\cite{zaccour2025access}. 

\paragraph{Synthetic Datasets} Synthetic datasets offer an alternative approach to fairness monitoring by statistically reproducing real-world data patterns, enabling analysis without relying on actual user data~\cite{stadler2022search,pereira2024assessment}. However, their reliability remains a key concern, as unavoidable discrepancies between synthetic and actual user data may undermine the validity of fairness assessments~\cite{zaccour2025access}. Synthetic data also presents some ethical challenges, notably those related to consent and ``diversity-washing''~\cite{whitney2024real}. In the context of fairness, synthetic data is viewed as a way of introducing more diversity into a dataset, but this synthetic data is still grounded in the biases of the original dataset, and can lead to biases~\cite{whitney2024real}.

\paragraph{Multi-Party Computation (MPC)} MPC offers another promising approach to privacy-preserving fairness monitoring by enabling multiple parties to jointly compute fairness metrics without revealing the raw sensitive data to any party\modtext{~\cite{helminger2022multi,espiritu2024synq,qin2019usability}. For example, the Boston Women's Workforce Council applies MPC to measure wage gaps across gender and racial groups without requiring any parties to reveal private information~\cite{lapets2018accessible,BWWC,qin2019usability}.} A common instantiation is secure two-party computation, in which a trusted third party collaborates with the model owner to compute fairness metrics using encrypted data from each side~\cite{helminger2022multi,de2023application,pinkas2009secure}. 

\paragraph{Our Contributions.} In contrast to DP and synthetic datasets, which rely on noise injection or distributional approximations, MPC maintains higher fidelity by enabling fairness metrics to be computed directly on encrypted data while still providing strong privacy guarantees. This advantage, however, comes with practical challenges. For instance, ensuring that individuals feel comfortable with distributed data storage involving a trusted third party, and determining who should serve as such a party, are open questions. To address these issues, we adopt a human-centered approach to study user acceptance of MPC protocols under varying design conditions, aiming to inform protocol designs that reconcile technical rigor with user expectations of fairness and privacy.

% Though it provides strong privacy guarantees, the practical application of two-party computation for fairness monitoring remains challenging especially considering user trust and acceptance, e.g., whether the mechanism of distributed data storage by involving a trusted third party could get user trust, and how to select such a party without discouraging data sharing. To address this gap, we took a human-centered approach to examine user acceptance of MPC protocol for fairness monitoring when varying the design attributes, aiming to enlighten user-friendly protocol design that considers user expectations of fairness and privacy.

% More importantly, when applying two-party computation for fairness monitoring, how to design this privacy-preserving fairness monitoring protocol that considers user expectations of fairness and privacy.

\subsection{User Acceptance of Data Sharing under Privacy Constraints}

Data donation, where individuals voluntarily contribute their personal data for public interest purposes, has become an important approach to data collection for societal benefit projects~\cite{gomez2023beyond,walquist2025collective}, including the development of fair algorithms~\cite{fabris2022algorithmic}. For instance, researchers in the HCI community adopted data donation to investigate user engagement in short videos~\cite{zannettou2024analyzing}, detect adolescent online risks~\cite{razi2022instagram}, and support trauma-informed design~\cite{zheng2024s}. 

However, despite the progress in privacy-preserving protocols for user-shared data~\cite{franzen2024communicating,helminger2022multi}, existing data donation efforts often emphasize the openness of data sharing, while inadequately informing volunteers about how their data will be used~\cite{gomez2023beyond} or protected~\cite{bowser2017accounting,franzen2024communicating}. Understanding user acceptance regarding data sharing under privacy-preserving protocols not only contributes to the goals of \textit{transparency} and \textit{meaningful consent} of data donation, but also provides valuable insight into \textit{protocol design} that reflects users' expectations~\cite{franzen2024communicating,xiong2020towards,kuhtreiber2022replication,cummings2021need}.

% Moreover, though recent research has increasingly highlighted privacy-preserving computational approaches for user-donated data (e.g., multi-party computation~\cite{helminger2022multi} and differential privacy~\cite{franzen2024communicating}), the transparency of these protocols to users, as well as how users establish trust and acceptance regarding data donation, remain underexplored.

Toward this goal, researchers have broadly investigated the technical and human factors that may influence user acceptance of data sharing under privacy constraints~\cite{kmetty2024determinants,franzen2024communicating,richter2021secondary,wenz2019willingness}. One line of work examined the effects of \textit{incentives} on data sharing, including monetary incentives~\cite{silber2022linking}, personal information gains~\cite{bietz2019data}, and contributions to social good~\cite{skatova2014data}. For example, Bietz et al. found that offering summary reports based on shared data could motivate participation in medical web surveys~\cite{bietz2019data}. \textit{Data type}, as a factor inherently relevant to privacy risks, has also substantially influenced users' willingness to share data (e.g., social media posts vs. search data~\cite{pfiffner2023leveraging}, photos and videos vs. less sensitive modalities~\cite{kmetty2024determinants}, GPS location vs. less confidential smartphone use~\cite{wenz2019willingness}). Recent studies have also begun to focus on the role of \textit{privacy protection mechanisms} and corresponding privacy communications in shaping users' data sharing decisions\modtext{~\cite{franzen2024communicating,xiong2020towards,cummings2021need,nanayakkara2023chances}}. For instance, Franzen et al. investigated how privacy communications influence data sharing and highlighted their significance in informing users about privacy protection mechanisms and potential risks~\cite{franzen2024communicating}. \modtext{Kacsmar et al. revealed that even though end users struggled with abstract definitions of private computation such as MPC, they found the concrete scenarios for private computation enlightening and plausible, which increased their acceptance of data sharing; nonetheless, the purpose of data sharing and analysis still served as the principal determinant of their attitudes~\cite{kacsmar2023comprehension}.} Finally, researchers have paid close attention to how \textit{human characteristics} (e.g., digital literacy~\cite{baumgartner2023novel}, privacy concerns~\cite{pfiffner2023leveraging}, and psychological traits~\cite{silber2022linking,kmetty2024determinants}) affect data sharing decisions. Based on these considerations, we detail how we contextualize MPC protocol design attributes that may influence user acceptance in Section~\ref{features}.
% \textit{Privacy} also substantially influences users' data donation decisions. 

\paragraph{Our Contributions.} Compared to other data donation scenarios, fairness monitoring relies on sensitive attributes (e.g., gender, race, and other demographic factors) to evaluate and mitigate algorithmic bias. The usage of sensitive attributes creates a context where users may need to weigh the benefits of contributing to algorithmic fairness against potential privacy risks. Therefore, understanding user preferences for protocol design is essential to developing user-trusted protocols that both enable fairness monitoring and safeguard privacy. This study aims to address this gap and propose practical implications for the design and communication of privacy-preserving fairness monitoring protocols.

% Existing research on user acceptance of data donation offers a solid foundation for exploring its use in fairness monitoring, a context in which users must balance the benefits of contributing to fairness against potential privacy losses. Yet, despite the breadth of prior work, few studies have systematically examined how specific design features of privacy-preserving protocols influence user perceptions, leaving a gap in developing a user-centered data-donation protocol. 

% little work  Privacy-preserving Protocols

% For specific donation systems, the perceived ease of use also affects users' willingness to donate data, e.g., the data upload time~\cite{kmetty2024determinants}.

\section{Situating the Study: MPC Protocols for Fairness Monitoring}\label{StudyContexts}

Real-world deployments of fairness monitoring occur in specific algorithmic contexts. In this paper, we situate our study in the domain of algorithmic hiring, an application that is particularly susceptible to discriminatory outcomes arising from historical and societal biases~\cite{fabris2024fairness, he2025developing} and thus designated as high-risk under the AI Act. 

In this paper, we assume multi-party computation protocols as a security and privacy measure safeguarding sensitive data to be processed for fairness monitoring. MPC allows for joint computation of fairness metrics using inputs from multiple parties without requiring any party to disclose the value of its inputs to preserve the confidentiality of individual sensitive attributes~\cite{goldreich1998secure,helminger2022multi}. The procedure of secure two-party computation, as an example of an MPC protocol for fairness monitoring, consists of the following steps:

\begin{enumerate}
    \item \textbf{Generation and deposit of two-party components}: Following the completion of the hiring process, a candidate may voluntarily share their sensitive data through a service managed by a trusted third party (TTP). The TTP then generates two-party components. Specifically, let \( x^{(\text{sens})}_i \in \mathbb{R} \) be the sensitive attribute (e.g., gender, race) of user \( i \) and \( r_i \sim \mathcal{R} \) be a random secret value generated by a trusted third party (TTP), \( c_{i,1} = x^{(\text{sens})}_i + r_i \), and \( c_{i,2} = x^{(\text{sens})}_i - r_i \) are generated as the secret shares. The two secret shares are then deposited to TTP and the hiring company independently; in this way, each user's sensitive attribute is secret-shared into two masked components, distributed to separate parties, and no party alone can infer \( x^{(\text{sens})}_i \).

    \item \textbf{Secure MPC for fairness computations}: During fairness evaluation, the sensitive attribute is virtually reconstructed under MPC:

    \[
    \langle x^{(\text{sens})}_i \rangle = \frac{c_{i,1} + c_{i,2}}{2}
    \]
    
    \noindent The value \( \langle x^{(\text{sens})}_i \rangle \) remains secret-shared and is used for the computation of fairness metrics. For example, given \( \langle x^{(\text{sens})}_i \rangle \), the mean outcome of a group $G$ (e.g., hiring or interview rate) is:
    \[
    \mu(G) = \frac{\sum_{i=1}^{N} y_i \cdot \mathbb{1}\!\left[ \langle x^{(\text{sens})}_i \rangle \in G \right]}
                   {\sum_{i=1}^{N} \mathbb{1}\!\left[ \langle x^{(\text{sens})}_i \rangle \in G \right]}.
    \]
    It can be further used to compute the demographic disparity~\cite{calders2010three} to measure algorithmic fairness, defined as the difference in mean group outcomes between groups:
    \[
    \delta = \mu(a) - \mu(b),
    \]
    All fairness computations are conducted over secret-shared inputs \( \langle x^{(\text{sens})}_i \rangle \). Therefore, at no point is any individual's sensitive attribute revealed to either party.
    
\end{enumerate}

% Compared to other privacy-preserving protocols for fairness monitoring such as synthetic datasets~\cite{stadler2022search,pereira2024assessment} and differential privacy~\cite{zaccour2025access,chang2021privacy,jagielski2019differentially}, MPC enables fairness monitoring without introducing noise or altering data distributions, thus preserving analytical accuracy by design. 

Though MPC protocols for fairness monitoring strike a reasonable balance between privacy guarantees for sensitive data and analytical accuracy in fairness evaluation, they also introduce several deployment complexities in real-world implementations~\cite{helminger2022multi,de2023application,lapets2018accessible}, such as selecting a trusted TTP the users would feel comfortable depositing their data with, communicating the protocol to enable meaningful consent, or fostering trust among different demographic groups to ensure that a diverse and representative sample of sensitive data can be collected for accurate fairness monitoring. \modtext{In fact, \textit{designing} privacy-preserving protocols from theory to practical implementation, and effectively \textit{communicating} these protocols to non-experts, remain key challenges in the real-world adoption of privacy-preserving technologies~\cite{agrawal2021exploring}. While recent work has explored experts' views~\cite{agrawal2021exploring} and end users' attitudes~\cite{kacsmar2023comprehension} toward the usability of privacy-preserving computation, including MPC, a quantitative analysis of user acceptance across different specific protocol designs could offer necessary insights for the effective deployment and communication of real-world protocols.}

% Recent studies have highlighted the importance of bridging the gap between privacy-preserving computation and real-world design, and communicating privacy-preserving computation

% 

To study this problem, we first derive the design dimensions of MPC protocols for fairness monitoring that relate to users' risk and benefit perceptions in Section~\ref{features}, and examine users' acceptance and preferences across different protocol designs (RQ1). Further, we investigate how users' fairness and privacy orientations and contexts shape their protocol design acceptance (RQ2), aiming to inform effective protocol communication to ensure informed consent across different populations.

% \section{Theoretical Foundations for Factors of User Trust}\label{features}

\section{From Protocol Designs to User Acceptance: A Privacy Calculus Perspective}\label{features}

% \subsection{RQ1: Features of Fairness Monitoring Protocol Design in Influencing Data Donation Willingness}
% Leveraging the ``5W1H'' guideline~\cite{gao2024taxonomy,jang2005unified}, we first aim to comprehensively capture the design attributes of the MPC protocol for fairness monitoring that may affect user perceptions, including \textit{fairness objective} (Why), \textit{collected information type} (What), \textit{data storage} (Where), \textit{monitoring actor} (Who), and \textit{monetary incentive}, \textit{data use} and \textit{privacy protection mechanism} (How). 
\modtext{Inspired by the ``5W1H'' guideline~\cite{gao2024taxonomy,jang2005unified}, we discuss how different design attributes of the MPC protocol for fairness monitoring presented in Section \ref{StudyContexts} may relate to user perceptions of data sharing, capturing aspects including \textit{fairness objective} (Why), \textit{collected information type} (What), \textit{data storage} (Where), \textit{monitoring actor} (Who), and \textit{monetary incentive}, \textit{data use} and \textit{privacy protection mechanism} (How). We then adopt Privacy Calculus Theory (PCT)~\cite{culnan1999information} to frame how these design attributes may influence individuals' willingness to share their data within the protocol.}
PCT posits that individuals weigh the \textit{perceived benefits} against the \textit{perceived risks} when deciding whether to share their information~\cite{culnan1999information}. It has been widely applied to understand how individuals make data sharing decisions in digital contexts~\cite{bemmann2024impact, wang2016intention, li2016examining}. Building on PCT, we categorized the design attributes of the MPC fairness monitoring protocol into benefit-related attributes (\textit{fairness objective} and \textit{monetary incentive}), risk-related attributes (\textit{privacy protection mechanism}, and \textit{data storage}), and attributes related to both (\textit{collected information type}, \textit{monitoring actor} and \textit{data use}).
% The design of MPC protocol for fairness monitoring . 

\subsection{Benefit-related Attributes}

\subsubsection{Fairness Objective: \textit{Which fairness objective does the data donation contribute to?}}\label{societalBenefit}
% data donation -> societal benefit

Data donation enables individuals to voluntarily share their personal data to support collective goals, such as advancing research or promoting social good~\cite{bietz2019data,skatova2014data}. Accordingly, the societal benefit, i.e., how data donation contributes to the public good, is one intrinsic incentive for potential data donors~\cite{skatova2014data,alashoor2025privacy}. Extensive work has suggested that understanding the purpose of data collection correlated with the stated willingness to share one's data (e.g.,~\cite{pfiffner2023leveraging,skatova2019psychology,sleigh2018experiences}).

% (e.g., process fairness in contrast with outcome fairness)

In the context of data donation for fairness monitoring, the core design attribute shaping the societal benefit is the selection of fairness objectives. Researchers in algorithmic fairness have developed different fairness objectives to evaluate and enhance algorithms~\cite{fabris2024fairness}. Different fairness objectives may correspond to distinct, or even conflicting, value assumptions~\cite{friedler2021possibility}. Among different fairness objectives~\cite{fabris2024fairness,caton2024fairness,mehrabi2021survey}, \textit{individual fairness} and \textit{group fairness} are two of the most widely adopted, despite their often conflicting goals~\cite{caton2024fairness,mehrabi2021survey}. Specifically, \textit{individual fairness} aims to give similar individuals similar decisions, while \textit{group fairness} emphasizes treating different groups equally~\cite{friedler2021possibility}. Different fairness objectives determine the direction in which algorithms are assessed and improved, and influence the perceived value and utility for potential data donors. Therefore, we investigate how different fairness objectives influence individuals' acceptance of fairness monitoring protocols.

\modtext{Notably, explaining algorithmic fairness to non-experts is challenging~\cite{shulner2022fairness,yan2024exploring}. Rather than focusing on mathematical definitions, we convey the intuitions behind fairness objectives to help users interpret optimization goals. Specifically, we use ``\textit{treat different demographic groups similarly}'' and ``\textit{treat similar individuals similarly}'' to capture group and individual fairness respectively, and include ``\textit{select the most qualified candidates}'' as a basic performance value for comparison. The inclusion of more nuanced fairness definitions extends previous work on public perceptions of fairness objectives (e.g., binary accuracy-fairness trade-offs~\cite{mourali2025public}). Although these objectives can overlap operationally (e.g., collected data can support both group and individual fairness monitoring), examining how users compare them offers insights into their fairness expectations. However, since the compatibility between fairness objectives may be interpreted by users differently, the corresponding findings should be viewed with caution - lower perceived importance does not imply lower practical value.}

 % - lower preference does not necessarily imply low practical value

% Therefore, we hypothesize that different fairness metrics can influence the decision-making of data donation for fairness monitoring (H1).
% different metrics -> fairness monitoring

% From the operational perspective of fairness monitoring, these values are not entirely mutually exclusive (e.g., the collected data can be used for monitoring both individual fairness and group fairness). However, examining how users compare different fairness objectives can enlighten different users' expectations of fairness optimizations.

\subsubsection{Monetary Incentive: \textit{How is the data donor compensated?}}

Financial incentives have become a topic of concern for data sharing as organizations seek to encourage greater participation~\cite{kmetty2024determinants,fast2023data}. Although monetary incentives may seem an obvious lever for increasing participation, the literature raises several concerns~\cite{fast2023data,silber2022linking}. For example, Fast et al. found that monetary incentives provided a short-term boost to data donation, but the effect was not sustained due to experienced opportunity costs and limited perceived benefits~\cite{fast2023data}. Silber et al. noted that high incentives can make respondents suspicious, leading them to assume their data is highly valuable, which may ultimately reduce willingness to participate~\cite{silber2022linking}. In addition, there are concerns that financial incentives might unduly influence or coerce individuals and undermine informed consent~\cite{ambuehl2015more}. Consistent with this, GDPR requires that consent be freely given, without coercion~\cite{hoofnagle2019european, EUdataregulations2018}. 

In this work, we include monetary incentives to examine how participants weigh their importance relative to other protocol design attributes. However, willingness to share data under monetary incentives should be interpreted cautiously and in light of ethical and legal considerations~\cite{berke2024insights}.
% , 
% , and hypothesize that monetary incentives can have a positive influence on participants' data donation (H2)

\subsection{Risk-related Attributes}

\subsubsection{Privacy Protection Mechanism: \textit{How is data protected?}}

Researchers have increasingly looked into privacy-preserving frameworks and technologies supporting data donation~\cite{boeschoten2022framework,appenzeller2022towards,helminger2022multi}. However, as privacy protection mechanisms are often perceived as too complex to keep participants adequately informed, only a few studies have examined their effects on individuals' decision-making for data sharing~\cite{franzen2024communicating,xiong2020towards,cummings2021need}. For example, recent work by Franzen et al. revealed that helping users understand the privacy protection mechanism (DP in their case) with a combination of graphical risk visualization and interactive risk exploration can support informed decision-making for data sharing~\cite{franzen2024communicating}. Xiong et al. found that showing descriptions that explain the implications, instead of the definitions/processes, could help participants better understand DP~\cite{xiong2020towards}.

Though MPC offers a technically strong privacy protection mechanism, how users trust and accept MPC to protect the shared data remains unclear, as MPC involves more stakeholders and more complex data processing. In this study, we investigate how the privacy protection mechanism of MPC influences user acceptance of the fairness monitoring protocol. Drawing insights from prior studies~\cite{franzen2024communicating,xiong2020towards}, we chose not to present technical definitions of MPC. Instead, we used visualizations to convey the intuitive foundations of how MPC enables privacy protection (\textit{distributed storage + encrypted computation}) to the study participants, as illustrated in Appendix~\ref{SurveyDetails}. That facilitates the comparison with the baseline condition in the survey study, i.e., \textit{anonymization and encryption}. 

% The MPC protocol for fairness monitoring is 

% Drawing insights from these studies, we used concise visualizations to introduce the intuitive foundations of how MPC enables privacy protection (distributed storage + encrypted computation) to the study participants. Though MPC offers a stronger privacy protection mechanism, it involves more stakeholders and more complex data processing that might hinder users' trust. In this work, we investigate whether privacy protection with MPC, in addition to the basic anonymization, would increase participants' willingness to donate data for fairness monitoring.
% Xiong et al. found that showing descriptions that explain the implications, instead of the definition/processes, could better support participants to comprehend DP~\cite{xiong2020towards}. 

% Our study situates the data-protection mechanism in multi-party computation (MPC), a representative privacy-protection scheme with GDPR compliance~\cite{helminger2022multi}. 

% Briefly speaking, under the MPC protocol for fairness monitoring, the data is split into encrypted pieces and shared between different parties, and different parties jointly perform computations for fairness metrics on encrypted data without either party revealing their private data. 

\subsubsection{Data Storage: \textit{Where is the encrypted data stored?}}

Recent work suggests that the entity storing data influences participants' willingness to share sensitive data~\cite{richter2021secondary,hillebrand2023social}. Generally, the perceived security and reputation of the data storage location play a key role in shaping trust. For instance, a survey study in Germany revealed that participants favored storing their donated data in a nationwide centralized database over long-term storage at research institutions~\cite{richter2021secondary}. An online experiment in the United States also suggested that people preferred their data managed by academic or government institutions over private entities~\cite{hillebrand2023social}.

The MPC protocol assumes a separate and independent entity, distinct from the service provider operating the job application system, as the trusted third party (TTP) that supports distributed storage of sensitive data and encrypted fairness computations. Therefore, it requires both technical capacities with necessary infrastructure, as well as a commitment to protecting data donors' privacy~\cite{helminger2022multi,ashurst2023fairness}. Ensuring users' trust in the TTP is essential for real-world deployment of the MPC protocol.

Therefore, we examine how users prioritize data storage locations relative to other protocol design attributes, and how they perceive, trust, and accept different TTPs as data storage sites, including \textit{research centers}, \textit{non-governmental non-profit organizations}, and \textit{specialized commercial companies}.

% We hypothesize that different data storage parties influence participants' willingness to donate data for fairness monitoring (H5).

% , and has gained increasing attention for research investigating what value individuals place on the utility and privacy in the digital era

\subsection{Attributes Related to Both Perceived Benefits and Risks}

\subsubsection{Collected Information Type: \textit{Which data is collected?}}

The type and scope of data requested shape how individuals perceive the privacy risks and data donation values, further influencing data-donation decisions~\cite{kmetty2024determinants,silber2022linking,breuer2023user,wenz2019willingness}. In particular, the perceived identifiability and sensitivity of collected data are significant factors influencing individuals' willingness to share data~\cite{kmetty2024determinants}. For example, Wenz et al. found that participants were less willing to share smartphone GPS data with geolocation compared to the accelerometer or questionnaire data~\cite{wenz2019willingness}. Similarly, Silber et al. uncovered that health app data had much lower overall sharing rates (6.1\%) compared to social media data such as Twitter (24\%)~\cite{silber2022linking}.

Situating the study in fairness monitoring for algorithmic hiring, we categorize the collected data types into three possible conditions:

\begin{enumerate}
    \item \textit{Demographic information} (e.g., age and gender), which constitutes the minimal data needed to measure input fairness~\cite{rastegarpanah2020fair}, reflected in whether the composition of the candidate pool is fair and representative;

    \item \textit{Demographic information + hiring decisions} (e.g., interview invitations or job offers), which enable richer monitoring for outcome fairness such as demographic parity (equal representation of different demographic groups in hiring system outcomes~\cite{caton2024fairness}); and

    \item \textit{Demographic information + hiring decisions + qualifications} (e.g., educational background and work experience), which support more complex monitoring across output fairness (e.g., equal exposure for different groups in the ranked results~\cite{fabris2024fairness}) and outcome fairness (e.g., equal opportunity for equally qualified individuals across groups~\cite{caton2024fairness,fabris2024fairness}) that require model input.
\end{enumerate}

These three conditions support fairness-monitoring tasks of varying complexity and produce outcomes of differing richness, each associated with a distinct level of privacy risks and perceived usefulness. In this study, we examine how changes in the collected data type influence participants' acceptance of the fairness monitoring protocol based on the three conditions.

\subsubsection{Monitoring Actor: \textit{Who is collecting data for fairness monitoring?}}

The stakeholders managing the data donation project for fairness monitoring also significantly influence willingness to share data. With regard to perceived benefits, individuals hold differing levels of trust in different organizations to collect data for societal good~\cite{liu2017donating}. In terms of perceived risks, individuals express less concern about privacy violations for organizations with a positive reputation, such as academic institutes compared to private industry~\cite{hillebrand2023social,liu2017donating}.

In our work, we hypothesize two scenarios of data donation with different monitoring actors: (1) data donation organized by \textit{auditing agencies and regulators} (external stakeholders); and (2) data donation organized by \textit{companies developing the hiring algorithms} (internal stakeholders). Through our survey, we evaluate how different stakeholders managing the data donation project affect participants' acceptance of the protocol.

\subsubsection{Data Use: \textit{What is data used for?}}

% The influence of the purpose of data use on participants' willingness to donate has been widely documented~\cite{pfiffner2023leveraging,skatova2019psychology,sleigh2018experiences}, and is also covered in Section \ref{societalBenefit}. In this work, the setting of data donation for fairness monitoring has necessitated the understanding of another important feature of data use: the data use approach, as detailed below.

According to Article 10(5) of the EU AI Act~\cite{EU_AI_Act_Article_10}, AI providers ``\textit{may exceptionally process special categories of personal data, subject to appropriate safeguards for the fundamental rights and freedoms of natural persons}'' to ensure bias detection and correction in relation to high-risk AI systems. The exception only applies to ``\textit{training, validation, and testing data sets}''~\cite{EU_AI_Act_Article_10}. Although such legal regulations define the conditions and extent of data use, how the public accepts varying levels of data use for fairness monitoring remains a significant concern. Indeed, data use may influence both the perceived benefits and risks of data donation. When users' data is employed solely for fairness evaluation to detect biased or discriminatory behaviors, the privacy risks are relatively low, as there is no direct influence on the algorithm. However, it requires additional fairness interventions to improve algorithmic fairness, such as using synthetic datasets, which may compromise perceived utility~\cite{liu2025can,cheng2021can}. On the other hand, training fair algorithms with user-shared data, in addition to evaluation, can have a more direct impact on algorithms but raises heightened security and privacy concerns, including the potential exposure of sensitive information via training-data inference attacks~\cite{menard2024artificial}. 

Therefore, we propose two scenarios in our survey: (1) \textit{data used only for model evaluation}, and (2) \textit{data used for model evaluation and development}. We examine how different modes of data use influence participants' willingness to share data under the protocol.

% On the one hand, the basic data use in fairness monitoring, i.e., evaluating the algorithms to surface the fairness-related issues, . 

% In the data donation for fairness monitoring, the data use. 

% We predict that different stakeholders managing the data donation project can affect participants' willingness to donate data (H7).

% \subsection{RQ2: Human and Environmental Features in Influencing Data Donation Willingness}

% In addition to technology factors (the design of fairness monitoring protocol), human and environmental factors may also influence users' willingness to donate data for fairness monitoring. In this section, we introduce the survey design to understand how human and environmental factors correlate with (1) participants' donation willingness and (2) participants' considerations regarding various protocol designs.

\section{Method}

In this section, we present the survey instrument for examining RQ1 and RQ2. For RQ1, we conduct (a) a scenario-based conjoint analysis and (b) a direct attribute-ranking task, to assess user-perceived importance of protocol attributes, along with attribute-level ratings to measure user preferences for different attribute levels in Section~\ref{RQ1-method}. For RQ2, we apply regression analysis to capture how users' fairness and privacy orientations and contexts correlate with their acceptance of the protocol design in Section~\ref{RQ2-method}. Finally, we outline the final study design with the details of participant recruitment and survey flow in Section~\ref{StudyDesign}. The recruitment and study procedures were approved by our Institutional Review Board (IRB).

\subsection{RQ1: Understanding User Priorities and Preferences Regarding Different Designs of the MPC Protocol for Fairness Monitoring}\label{RQ1-method}

To investigate user priorities and preferences of various designs of MPC fairness monitoring protocols (RQ1), we first explore how users weigh different design attributes with two complementary approaches: a scenario-based conjoint analysis to measure \emph{revealed} attribute importance, and a direct attribute ranking to measure \emph{stated} attribute importance. In addition, we include attribute-level rating questions to evaluate users' preferences for specific attribute levels in greater detail.

 % aim to examine the role of different design attributes in shaping user acceptance of the MPC protocol for fairness monitoring. Therefore, we measure the attribute importance

% To explore user acceptance of various designs of MPC fairness monitoring protocols (RQ1), we first measure user-perceived attribute importance with two complementary approaches: choice-based conjoint analysis to measure revealed attribute importance, and direct attribute ranking to measure stated attribute importance. 

\subsubsection{Conjoint Analysis to Measure Revealed Attribute Importance}

We first adopt conjoint analysis to investigate how different design attributes of the MPC protocol for fairness monitoring influence users' decision-making for data sharing~\cite{green1971conjoint}. Conjoint analysis is a statistical technique commonly employed to examine how individuals make complex decisions by evaluating trade-offs among various attributes. It is most widely applied in the field of marketing and has gained growing prominence in HCI and usable privacy research exploring the design space of digital applications~\cite{ayalon2023exploring,pu2017valuating,ladak2024artificial}. Through a series of hypothetical scenario-based choices of different protocol designs, we measure how users weigh different design attributes, as summarized in Table~\ref{CBCAttributes}, based on the discussion in  Section~\ref{features}. This approach facilitates comparative evaluation of design alternatives at scale, allowing us to unpack how users navigate the trade-offs inherent in data sharing, particularly between contributing to fairness and protecting privacy.

\begin{table}[ht] 
    
    \centering
    \caption{Attributes Presented in the Choice-Based Conjoint Analysis Surveys.}
    \label{CBCAttributes}
    \begin{tabular}{|p{1.8cm}|p{2.5cm}|p{3cm}|p{6cm}|}
        \hline
        \textbf{Category} & \textbf{Attribute} & \textbf{Description} & \textbf{Possible Values} \\
        \hline
        \multirow{1}{3cm}{Benefit-related Attributes} & Fairness Objective & \textit{Which fairness objective does the data donation contribute to?} & (1) Select the most qualified candidates; (2) Treat different demographic groups similarly; (3) Treat similar individuals similarly \\

        \cline{2-4}
        & Monetary Incentive & \textit{How is the data donor compensated?} & (1) No monetary compensation; (2) \$1 gift card; (3) \$5 gift card \\

        \hline

        \multirow{1}{3cm}{Risk-related\\ Attributes}
        & Privacy Protection Mechanism & \textit{How is data protected?} & (1) Anonymize and encrypt the personal data; (2) Anonymize and encrypt personal data, then distribute the encrypted data so no single party has full access to sensitive personal data.\\

        \cline{2-4}
        & Data Storage & \textit{Where is the encrypted data stored?} & (1) Government; (2) Research centers; (3) Non-governmental non-profit organizations; (4) Specialized commercial companies \\

        \hline

        \multirow{1}{3cm}{Attributes\\ Related to\\ Both} & 
        Collected Information Type & \textit{Which data is collected?} & (1) Demographic attributes; (2) Demographic attributes and hiring decisions; (3) Demographic attributes, qualifications, and hiring decisions \\
        
        \cline{2-4} & Monitoring Stakeholder & \textit{Who is collecting data for fairness monitoring?} & (1) Auditing agencies and regulators; (2) Companies developing automated hiring systems \\

        \cline{2-4}
        & Data Use & \textit{What is data used for?} &  (1) Evaluate automated hiring systems; (2) Develop and evaluate automated hiring systems \\

        \hline

    \end{tabular}
\end{table}

% \subsubsection{RQ1: How do various designs of fairness monitoring protocols influence user acceptance and trust?}

% First, we used a \textbf{conjoint analysis} to assess how participants value different attributes of fairness monitoring protocols. 

\begin{figure}[h]
    \caption{Example of a Conjoint Task Designed to Examine Users' Trade-offs Among Protocol Design Attributes. Respondents choose between two MPC fairness monitoring protocol designs or ``none of these''. Conjoint analysis is applied to analyze their selections to reveal attribute importance.}
    \includegraphics[width=0.8\textwidth]{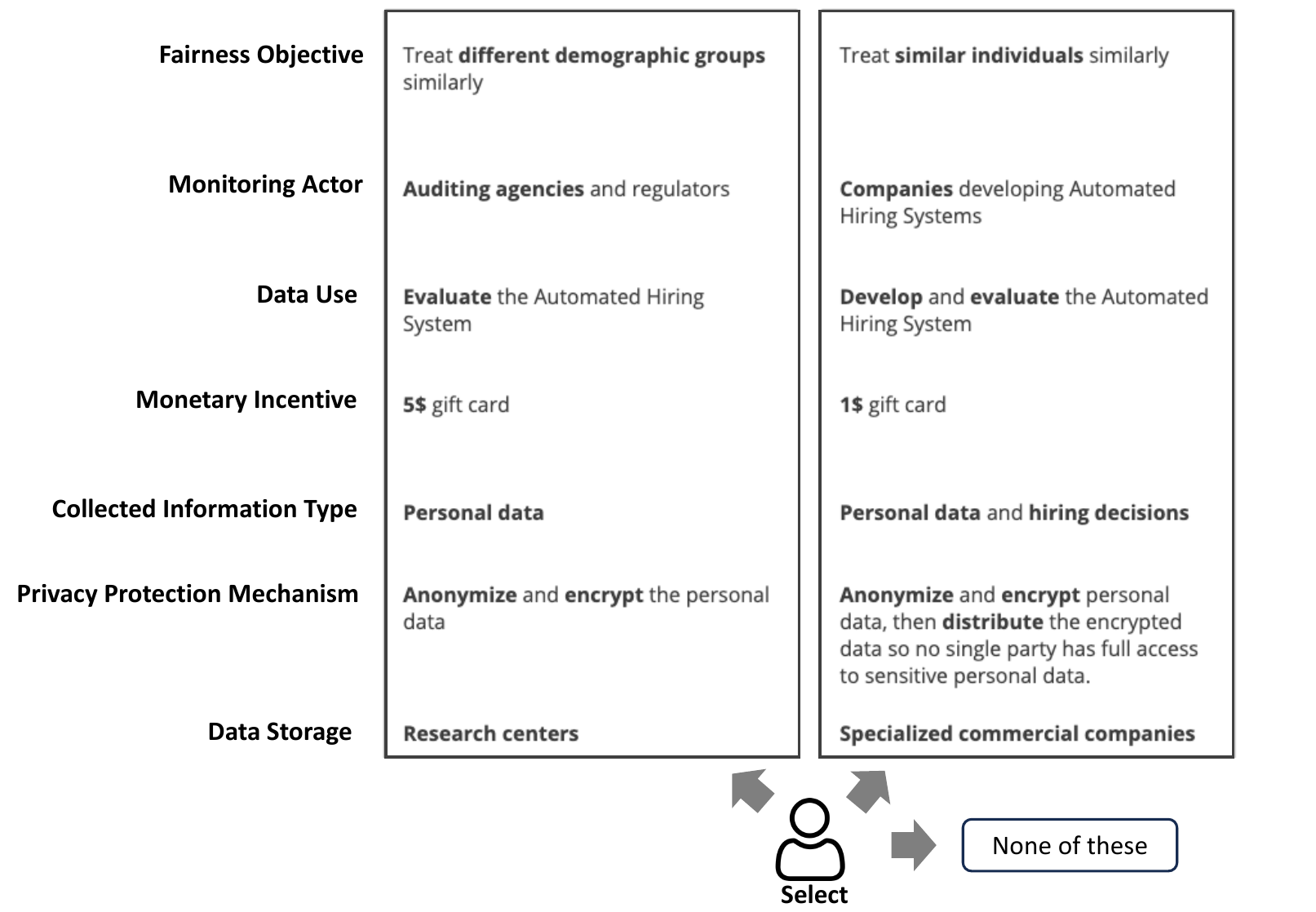}
    \label{fig:conjoint}
\end{figure}

Each participant was presented with five conjoint tasks, in which they were asked to select their preferred protocol designs. Each task included two design options across the seven attributes, along with a ``none of these'' option to mitigate forced-choice bias~\cite{green1971conjoint}. An example task is shown in Figure \ref{fig:conjoint}. The combination of attributes and their levels resulted in 864 possible protocol design versions (3 × 3 × 2 × 4 × 3 × 2 × 2). To manage this complexity and reduce the number of versions shown to participants, we used Sawtooth Software\footnote{https://sawtoothsoftware.com/} to generate an orthogonalized design, resulting in 300 unique protocol design versions~\cite{ayalon2023exploring}. 

Based on participants' choices, we estimated individual-level \textit{part-worth utilities} using the \textit{Hierarchical Bayes (HB)} model implemented in Sawtooth Software. This model assumes that respondents' choices follow a \textit{multinomial logit model}, where the probability of selecting an option is determined by the relative utility of that option compared to the others. Detailed explanations of utility estimation are provided in Appendix \ref{Utility-Estimation}. Attribute importance was then derived by calculating the relative range of part-worth utilities within each attribute, normalized across all attributes to sum to 100\%.

% \begin{comment}
% The utility for participant $i$ choosing option $j$ in task $t$ is defined as:

% \[
% P_{ijt} = \frac{e^{U_{ijt}}}{\sum_{k \in J} e^{U_{ikt}}}
% \]

% where the utility $U_{ijt}$ is defined as an additive function:

% \[
% U_{ijt} = \beta_0 + \sum_a \beta_a x_{ajt} + \epsilon_{ijt}
% \]

% Here, $\beta_a$ represents the utility weight of attribute $a$, $x_{ajt}$ indicates the level of attribute $a$ for option $j$ in task $t$, and $\epsilon_{ijt}$ captures unobserved variance. We calculated \textit{zero-centered utilities} for each respondent (summing to zero across attribute levels) and \textit{attribute importance scores}, which sum to 100\% per respondent. The relative importance of each attribute was computed as:

% \[
% I_a = \frac{\max(U_a) - \min(U_a)}{\sum_b (\max(U_b) - \min(U_b))}
% \]

% This method allowed us to identify how different design attributes influenced users' decision-making regarding the protocol design.
% \end{comment}

\subsubsection{Direct Attribute Ranking to Measure Stated Attribute Importance}\label{AttributeRanking}

In addition to revealed attribute importance inferred from conjoint analysis, we also measure stated attribute importance, allowing users to directly indicate their design priorities in an \textit{attribute-ranking task}: the survey directly presented the seven attributes and participants ranked the attributes from most to least important. We applied reversed scoring for importance estimation where higher numerical values indicate higher perceived importance. This direct ranking complements the conjoint tasks~\cite{eggers2009hybrid}; ranking offers a simple, intuitive expression of perceived importance, whereas choice-based conjoint prompts realistic trade-offs akin to real-world decision-making. 

% The collection of attribute importance through different cognitive mechanisms provides a more in-depth understanding of users' trade-offs of different protocol designs.

\subsubsection{Attribute-level Rating to Measure Attribute Design Preference}\label{AttributeRating}

Following the investigation of how users weigh different design attributes, we examine user preferences for specific design levels within each attribute. Specifically, we designed 7-point Likert-scale questions for every level of the seven attributes, allowing participants to rate their acceptance of each attribute level on a scale from -3 (least acceptable) to +3 (most acceptable). After collecting the data, we compared differences across levels within each attribute using Welch's ANOVA test~\cite{liu2015comparing}, as users' ratings were approximately normally distributed but showed unequal variances.

% To measure \textbf{stated attribute importance}, we designed a \textit{ranking question}, where the survey directly presented the seven attributes and participants were asked to rank the attributes from most to least important. We applied reversed ranking for importance estimation so that higher numerical values indicate higher perceived importance. The collection of attribute importance through different cognitive mechanisms provides a more in-depth understanding of users' trade-offs of different protocol designs.

\subsection{RQ2: Exploring How Users' Fairness and Privacy Orientations and Contexts Correlate with Protocol Design Acceptance}\label{RQ2-method}

After understanding user priorities and preferences regarding MPC protocol design, we further investigate how users' fairness and privacy orientations and contexts correlate with their acceptance of the protocol design, based on a regression analysis. This investigation can help inform the design of more inclusive protocols, as well as enable customized protocol communication for meaningfully informed consent.
% We adopt regression analysis to investigate how human factors correlate with acceptance of the protocol design, with the goals of informing more inclusive protocol design as well as enabling customized protocol communication for meaningfully informed consent.

\subsubsection{Factor Development}

% operationalize \textit{user orientations and contexts} that plausibly shape preferences over MPC protocol attributes for fairness monitoring, instead of using the broad notion of “human factors.”

Guided by the privacy calculus perspective, we operationalize users' fairness and privacy orientations and contexts as factors potentially shaping acceptance of MPC protocols for fairness monitoring, given the substantial role of user orientations and contexts in privacy calculus~\cite{seberger2021us,ghaiumy2024personalizing,woodruff2014would}. Specifically, we include \textit{Privacy Orientation} for users' stance toward privacy risks and \textit{Fairness Orientation} for users' valuation of fairness monitoring, and \textit{Demographics}, \textit{Prior Experience} and \textit{System Literacy} as contextual factors influencing fairness and privacy perceptions~\cite{wang2020factors,zukowski2007examining}. The study encompasses the following factors:

% For  Besides, We also included \textit{demographics}, \textit{user context} and \textit{system literacy} as common factors related to users' fairness and privacy perceptions~\cite{wang2020factors,zukowski2007examining}. 

% , , \textit{perceived importance of transparency of data processing}, and \textit{perceived importance of transparency of data protection}

\begin{itemize}
    \item \textbf{Privacy Orientation}: users' stance toward privacy risks and controls, measured by \textit{general data donation willingness}, \textit{perceived risks of data donation}, \textit{active privacy safeguards}, \textit{understanding of online privacy risks}, \textit{understanding of online data use}, \textit{perceived importance of data use transparency}, and \textit{perceived importance of data protection transparency}.

    \item \textbf{Fairness Orientation}: users' valuation of fairness monitoring, captured by \textit{perceived effectiveness of fairness monitoring systems} and \textit{perceived fairness of algorithmic hiring systems}.
    
    \item \textbf{System Literacy}: users' understanding and efficacy related to the systems, including \textit{understanding of automated hiring systems} and \textit{understanding of fairness monitoring systems}.

    \item \textbf{Prior Experience}: \textit{personal discrimination experience in any context}, \textit{personal discrimination experience in job applications}, \textit{perceived discrimination experience of close relations}, and \textit{prior experience with data donations}.

    \item \textbf{Demographics}: \textit{age}, \textit{gender}, \textit{ethnicity}, \textit{educational background}, and \textit{employment status}.

    % \item \textbf{(Optional) Transparency \& Governance Expectations}: if modeled separately from Privacy Orientation, the two transparency items serve as proxies for governance expectations regarding data processing and protection.
\end{itemize}

\subsubsection{Analysis}

We conducted linear regression analysis to investigate how users' fairness and privacy orientations and contexts correlated with user acceptance of the protocol (RQ2). Specifically, we performed separate regressions for each of the seven protocol design attributes using two different outcome measures: (1) the revealed importance estimated via the Hierarchical Bayes (HB) model from the conjoint task, and (2) the stated importance obtained from the ranking task. This resulted in a total of 14 regression models (7 attributes $\times$ 2 outcome types). We standardized the numeric features, and encoded categorical variables using one-hot representations. We applied L1 regularization to select features by shrinking less important coefficients to zero, reducing overfitting and improving interpretability.

\subsection{Study Design}\label{StudyDesign}

In this section, we outline the details of the study design, including the survey flow and participant recruitment.

% We conducted a survey study with a large and diverse sample to understand user acceptance of the MPC protocol design for fairness monitoring. This section outlines the survey structure and study procedures designed to address our research questions. We contextualized algorithmic fairness within \textit{algorithmic hiring}, a representative high-stakes domain that is particularly vulnerable to discriminatory outcomes arising from historical and societal biases~\cite{fabris2024fairness}.

% This approach facilitates comparative evaluation of design alternatives at scale, allowing us to examine how users navigate the trade-offs inherent in data-donation protocols, particularly between contributing to fairness and protecting privacy.

% \paragraph{Survey Design and Refinement.}

% % To ensure the survey accurately captures user perspectives on protocol design and acceptance, the authors first developed key dimensions based on relevant literature and technical implementation. These dimensions connect protocol design features to established factors influencing user acceptance.

% \paragraph{Participant Recruitment.}

% \paragraph{Data Analysis}

% We finally conducted data analysis across conjoint analysis, statistical tests, and regression analysis.

\subsubsection{Survey Flow}

Based on the theoretical foundations in Section \ref{features}, three authors collaboratively and iteratively constructed the survey, resolving disagreements through several rounds of discussions. A pilot study was then conducted with eight participants in the EU representing diverse cultural backgrounds (Europe, North America, and Asia) and varying levels of familiarity with privacy-enhancing technologies and algorithmic fairness. Feedback from this pilot study helped refine the survey by clarifying confusing elements and ensuring accessibility across knowledge levels. The full survey is detailed in Appendix \ref{SurveyDetails}.

\paragraph{Background Knowledge Introduction.}

To facilitate participants' understanding of the study, the survey began with an introduction to key background concepts, including automated hiring systems, fairness in such systems, the role of data donation in fairness monitoring, and the basic idea of MPC for protecting sensitive data. To reduce cognitive load, we employed visual examples (shown in Appendix \ref{SurveyDetails}) to illustrate: (1) how fairness issues emerge in automated hiring systems due to biased training data, (2) how data donation can aid in detecting such issues, and (3) how fairness monitoring can be conducted without revealing sensitive information through TTP-assisted distributed storage combined with MPC. To ensure participants had sufficient comprehension of the background material, we included three binary-choice assessment questions. Participants who did not pass the assessment were redirected to the introduction page and needed to answer all three questions correctly before proceeding to the survey. 

% The specific design for background knowledge introduction and assessment is detailed in Appendix \ref{Background}.

\paragraph{Collection of Users' Fairness and Privacy Orientations and Contexts.}

% The collection of user characteristics not only aimed to capture participants' backgrounds to facilitate the interpretation of findings, but also served to investigate how human factors correlate with user acceptance of MPC fairness monitoring protocol (RQ2).

After equipping participants with the necessary background knowledge, we collected users' fairness and privacy orientations and contexts as detailed in Section \ref{RQ2-method}, including \textit{demographics}, \textit{prior experience}, \textit{system literacy}, \textit{fairness orientation}, and \textit{privacy orientation}. This section also included an attention-check question; participants who failed to answer it correctly were excluded from the analysis.

\paragraph{Conjoint Tasks.}

Next, participants were presented with five conjoint tasks to indicate their preferences for protocol design as detailed in Section \ref{RQ1-method}. Based on pilot study feedback, participants found the conjoint tasks intuitive and easy for comparing attributes, so we kept attribute order consistent across tasks to avoid confusion given the repeated tasks; the results confirmed that attribute order had no significant effect on revealed importance. We also placed attention checks in other parts of the survey instead of in the conjoint tasks, since most attributes lacked a clearly dominant level over others (e.g., more monetary incentive might amplify the perceived risks; less collected data might limit users' perceived usefulness of data donation). Therefore, we could not design one option as unambiguously superior for attention check~\cite{ayalon2023exploring}.

% Each task included two design options across the seven attributes introduced in Section \ref{features}, along with a ``none of these'' option to mitigate forced-choice bias~\cite{green1971conjoint}. One example of the conjoint task is shown in Figure \ref{fig:conjoint}. The combination of attributes and their levels resulted in 864 possible protocol design versions (3 × 3 × 3 × 2 × 4 × 2 × 2). To manage this complexity and reduce the number of versions shown to participants, we used Sawtooth Software\footnote{https://sawtoothsoftware.com/} to generate an orthogonalized design, resulting in 300 unique versions~\cite{ayalon2023exploring}. 

\paragraph{Self-Declared Attribute Importance and Level Acceptance}

Following the conjoint tasks, participants were asked to complete an attribute ranking task for stated attribute importance as shown in ~\ref{AttributeRanking}, and an attribute-level rating task to specify their preferred design options as shown in ~\ref{AttributeRating}. 

% Finally, participants responded to 7-point Likert-scale questions across all attribute levels of the seven attributes, rating their acceptance of each attribute level on a scale from -3 (least important) to +3 (most important).

% C allowing us to evaluate the alignment between stated and revealed preferences.

% These self-declared responses were used to compare participants' perceived importance of attributes and levels with the results obtained from the conjoint analysis, allowing us to evaluate the alignment between stated and revealed preferences.

\subsubsection{Participant Recruitment}

Before participant recruitment, we conducted a power analysis to determine the required sample size, indicating a minimum of 626 participants assuming five conjoint tasks per participant~\cite{schuessler2020power,freitag2020cjpowr}. Next, we recruited participants through the Prolific platform.\footnote{https://prolific.co/} Eligible participants should reside in European countries where the GDPR is in effect, and current job-seekers as the target group of the MPC protocol for fairness monitoring in the hiring setting. Given that fairness and privacy perceptions might differ across demographic groups, we recruited participants in batches and monitored the demographic distribution. As the initial recruitment resulted in a predominantly white sample, not fully representative of the EU's demographics, an additional 300 responses were deliberately collected from non-white participants to improve demographic diversity. A total of 833 individuals participated in our study. The survey was designed to take approximately 10 minutes and was compensated with €2.

% \subsection{Descriptive Statistics}\label{finding:DesStats}

% Because the non-controlled recruitment initially resulted in a predominantly white sample, an additional 300 responses were deliberately collected from non-white participants, leading to

\begin{figure*}[htbp]
    \centering
    \includegraphics[width=\textwidth]{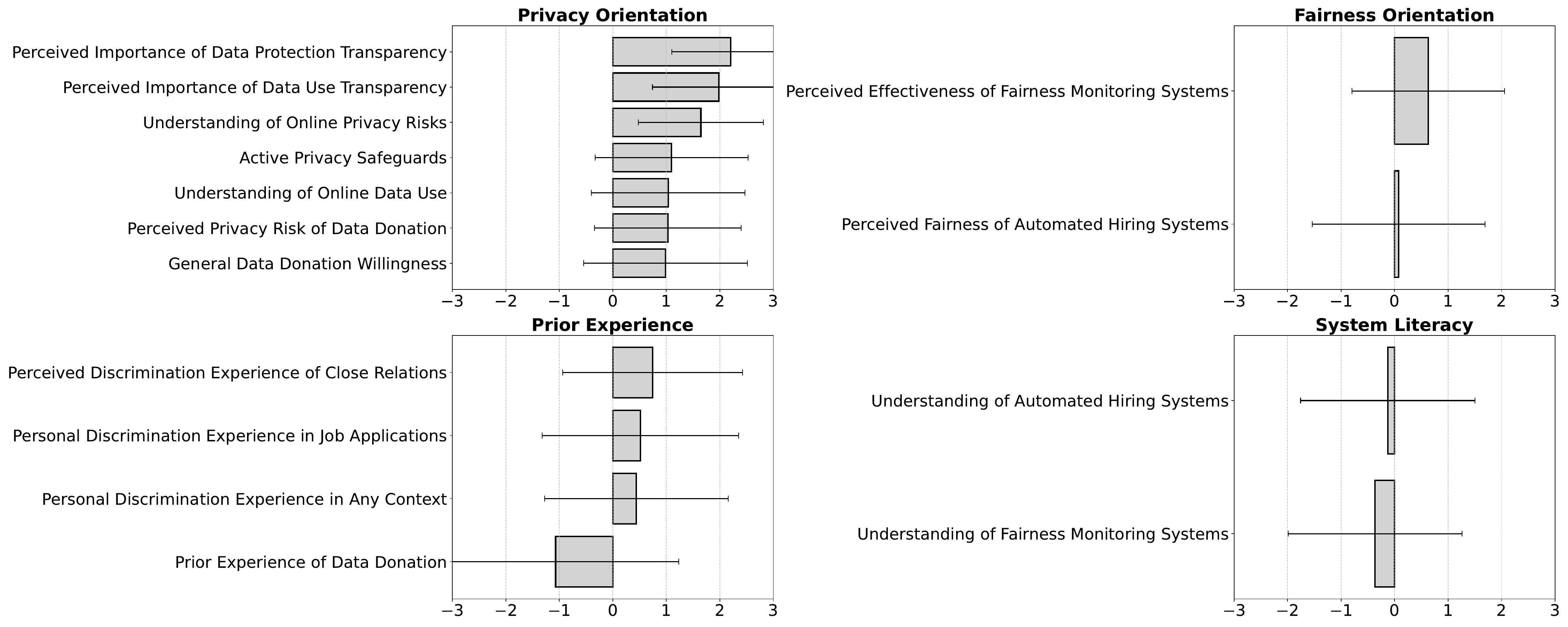}
    \caption{Means and Standard Deviations of Users' Fairness and Privacy Orientations, Prior Experience, and System Literacy}
    \label{fig:descriptive}
\end{figure*}

% The final dataset consisted of responses from 795 job-seekers, after excluding those who failed the attention check or completed the survey in under five minutes. 
After excluding participants who failed the attention check or completed the survey in less than 5 minutes, our final sample consisted of 795 participants. The sample was slightly \modtext{man}-dominated, with 53.5\% participants identifying as a \modtext{man} ($N = 425$), 45.0\% as a \modtext{woman} ($N = 358$), and 1.0\% as non-binary ($N = 8$). The sample has the following ethnic distribution: 60.4\% White ($N = 480$), 15.1\% Asian ($N = 120$), 15.0\% Black ($N = 119$), 7.9\% of mixed ethnic backgrounds ($N = 63$), and 1.5\% from other groups ($N = 12$). Among the participants, the largest age group was 25–34, comprising 39.9\% of the sample ($N = 317$), followed by the 18–24 group at 21.9\% ($N = 174$), and the 35–44 group at 20.8\% ($N = 165$). Respondents aged 45 and above represented a smaller portion of the sample, with 12.5\% aged 45–54 ($N = 99$), 4.2\% aged 55–64 ($N = 33$), and 0.9\% aged 65 or older ($N = 7$). Overall, 74.2\% of participants had attained university-level or higher education ($N = 590$), while 24.2\% had completed secondary education ($N = 192$), and 1.3\% had basic or no education ($N = 10$). Regarding employment status, 56.7\% of participants were employed ($N = 450$), 39.4\% were unemployed ($N = 313$), with the remaining not disclosing their status. The means and standard deviations of users' fairness and privacy orientations, prior experience, and system literacy are shown in Figure \ref{fig:descriptive}.

\section{Findings}

In this section, we introduce the findings of user acceptance of the MPC protocol for fairness monitoring. We illustrate users' priorities and preferences of different design attributes (RQ1) in Section \ref{RQ1-findings}, and unpack how users' fairness and privacy orientations and contexts correlate with user acceptance of protocol design (RQ2) in Section \ref{RQ2-findings}.

\subsection{RQ1: User Acceptance of Different MPC Fairness Monitoring Protocol Designs}\label{RQ1-findings}

\subsubsection{Attribute Importance in MPC Fairness Monitoring Protocol Designs}\label{RQ1-attributes}
% \begin{comment}
%     \begin{figure*}[htbp]
%     \centering
%     \includegraphics[width=0.95\textwidth, height=9cm]{MPC_pics/attribute_importances_plot.pdf}
%     \caption{Attribute Importance in Privacy-preserving Fairness Monitoring Protocol Design}
%     \label{fig:AttributeImportance}
% \end{figure*}
% \end{comment}
\begin{figure*}[htbp]
    \centering
    \includegraphics[width=0.95\textwidth, height=9cm]{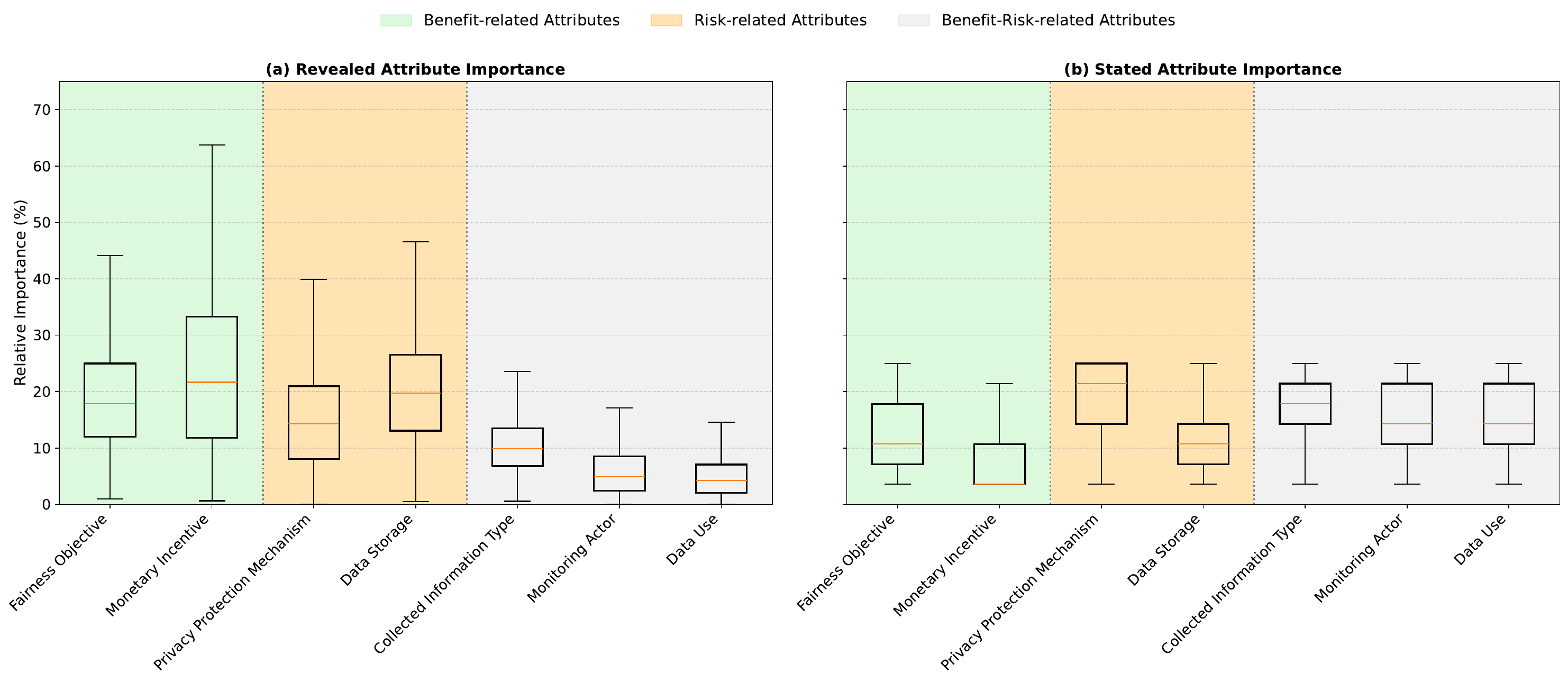}
    \caption{Attribute Importance in MPC Fairness Monitoring Protocol Design, Including (a) Revealed Importance Based on Choice-Based Conjoint Analysis and (b) Stated Importance Based on Attribute Ranking}
    \label{fig:AttributeImportance}
\end{figure*}

Figure \ref{fig:AttributeImportance} shows the attribute importance in the MPC protocol design, including (a) \textit{revealed} importance based on scenario-based conjoint analysis, and (b) \textit{stated} importance based on direct attribute ranking. 

% We found that users prioritized \textit{collected information type} and \textit{privacy protection mechanisms}, as factors related to privacy risks, in direct attribute rankings. However, in simulated decision-making scenarios, they paid greater attention to benefit-related factors, including \textit{the fairness objective} and \textit{monetary incentive}

\paragraph{Revealed attribute importance based on conjoint analysis: \modtext{participants prioritize benefit-related factors in simulated decision-making scenarios.}} Both benefit-related attributes, \textit{monetary incentive} and \textit{fairness objective}, strongly influenced users' decision-making for a preferred protocol design, ranking as the first (mean = 23.6\%, SD = 14.4\%) and third (mean = 19.2\%, SD = 9.8\%) most important attributes respectively. For risk-related attributes, \textit{data storage} was considered the second most important attribute (mean = 20.3\%, SD = 10.0\%), and \textit{privacy protection mechanism} had an average relative importance of 15.1\% (SD = 9.2\%). The three attributes concerning both benefits and privacy risks, \textit{collected information type}, \textit{monitoring actor} and \textit{data use},  had relative importance values of 10.8\% (SD = 5.8\%), 6.0\% (SD = 4.6\%) and 5.0\% (SD = 3.9\%) respectively.

\paragraph{Stated attribute importance based on direct ranking tasks: \modtext{participants prioritize collected information type and privacy protection mechanisms in direct ranking.}} Users' stated attribute importance, as captured by direct ranking, differed substantially from their revealed preferences in simulated conjoint tasks. In the ranking task, the most highly prioritized attribute was \textit{privacy protection mechanism} (mean = 19.1\%, SD = 5.4\%). \textit{Collected information type} ranked second in stated importance (mean = 17.6\%, SD = 6.0\%). In contrast, despite their significant influence on users' decisions in the simulated scenarios, the two benefit-related attributes, \textit{fairness objective} and \textit{monetary incentive}, were ranked among the lowest, with importance of 12.0\% (SD = 6.9\%) and 7.8\% (SD = 6.4\%).

\subsubsection{User Preferences for Different Attribute Levels in Protocol Design}\label{RQ1-attributeLevels}

\paragraph{General protocol design preferences.}

Table \ref{tab:acceptance_dimensions} presents user preferences for different attribute levels in protocol design. For \textit{fairness objective}, the general algorithm performance (i.e., select the most qualified candidates) outweighed group fairness (i.e., treat different demographic groups similarly) and individual fairness (i.e., treat similar individuals similarly), getting the highest acceptance score (Mean Acc. Score = 79.1\%, SD = 20.8\%). For \textit{monetary incentive}, higher monetary compensation correlated with a higher acceptance score.

Our findings indicate the potential openness among participants to consider partial privacy concessions when contributing to algorithmic fairness. For \textit{collected information type}, users had a higher acceptance score (Mean Acc. Score = 75.3\%, SD = 24.6\%) when the protocol collected all relevant information (including hiring decisions, qualifications and demographics), in comparison with only collecting basic demographic data. Similarly, users exhibited a higher willingness to contribute their data for developing and evaluating the automated hiring system (Mean Acc. Score = 78.5\%, SD = 19.2\%) compared to only evaluating the system (Mean Acc. Score = 72.0\%, SD = 19.6\%) regarding \textit{data use}.

In general, users placed greater trust in disinterested agencies than in commercial companies to serve as both the monitoring actor and the trusted third party (TTP) for data storage. Particularly, research centers received a higher acceptance score (Mean Acc. Score = 76.9\%, SD = 20.6\%) compared to non-governmental non-profit organizations (Mean Acc. Score = 72.8\%, SD = 23.3\%) and specialized commercial companies (Mean Acc. Score = 56.7\%, SD = 29.6\%) in serving as the TTP for \textit{data storage}. Besides, users expressed higher acceptance for auditing agencies and regulators (Mean Acc. Score = 73.6\%, SD = 21.8\%) as the \textit{monitoring actor} compared to companies developing automated hiring systems (Mean Acc. Score = 63.5\%, SD = 26.2\%). 

Finally, users viewed distributed storage, the MPC protocol's key \textit{privacy protection mechanism}, as more effective than basic anonymization and encryption. Though anonymization and encryption of personal data have achieved a high acceptance score (Mean Acc. Score = 76.2\%, SD = 19.5\%), adding distributed storage further increased acceptance by approximately 10\% (Mean Acc. Score = 86.5\%, SD = 18.7\%).

\begin{table}[htbp]
\small
\centering
\caption{User Preferences of Different Attribute Levels in Protocol Design. ``Mean Acc. Score'' presents the mean acceptance score for each attribute level among the participants, where 100\% represents ``most acceptable'' and 0 represents ``least acceptable''. The differences of all attributes are statistically significant under Welch's ANOVA test (*** p < 0.001)~\cite{liu2015comparing}, which is appropriate for our setting of approximately normal distributions with unequal variances.}
\begin{tabular}{|p{6.5cm}|p{1.7cm}|p{1.5cm}|}
\hline
\textbf{Attribute Level} & \textbf{Mean Acc. (\%)} & \textbf{Std. Dev. (\%)} \\
\hline
\multicolumn{3}{|c|}{\textbf{Fairness Objective***}} \\
% ($p\_value = 1.34*10 ^{-18} $)
\hline
Select the most qualified candidates & \textbf{79.10} & 20.80 \\
Treat different demographic groups similarly & 73.77 & 25.08 \\
Treat similar individuals similarly & 68.76 & 24.09 \\
\hline
\multicolumn{3}{|c|}{\textbf{Monetary Incentive***}} \\
% ($p\_value = 5.18 * 10^{-150}$)
\hline
No monetary compensation & 40.31 & 30.02 \\
\$1 gift card & 58.93 & 24.26 \\
\$5 gift card & \textbf{79.90} & 24.20 \\

\hline
\multicolumn{3}{|c|}{\textbf{Privacy Protection Mechanism***}} \\
\hline
Anonymization and encryption & 76.21 & 19.48 \\
Anonymization, encryption, and distributed storage & \textbf{86.48} & 18.72 \\
\hline
\multicolumn{3}{|c|}{\textbf{Data Storage***}} \\
\hline
Specialized commercial companies & 56.73 & 29.57 \\
Research centers & \textbf{76.86} & 20.60 \\
Non-governmental non-profit organizations & 72.83 & 23.27 \\
\hline
\multicolumn{3}{|c|}{\textbf{Collected Information Type***}} \\
\hline
Demographic data & 66.69 & 23.86 \\
Demographic data and hiring decisions & 70.23 & 21.39 \\
Demographic data, hiring decisions, and qualifications & \textbf{75.26} & 24.59 \\
\hline
\multicolumn{3}{|c|}{\textbf{Monitoring Actor***}} \\
\hline
Auditing agencies and regulators & \textbf{73.56} & 21.77 \\
Companies developing automated hiring systems & 63.52 & 26.24 \\
\hline
\multicolumn{3}{|c|}{\textbf{Data Use***}} \\
\hline
Evaluate automated hiring systems & 72.01 & 19.63 \\
Develop and evaluate automated hiring systems & \textbf{78.51} & 19.18 \\

\hline
\end{tabular}
\label{tab:acceptance_dimensions}
\end{table}

\paragraph{\modtext{Risk–benefit trade-offs vary by demographic groups.}}

\begin{table}[htbp]
\small
\centering
\caption{Comparison between Demographic Groups regarding User Acceptance of Different Protocol Design. We use ``His. Adv.'' to denote historically advantaged groups in labor market (younger/\modtext{man}/white job seekers), and ``His. Disadv.'' to denote historically disadvantaged groups as the counterpart. We only present attribute levels with statistically significant differences ($p \le 0.05$) between groups under t-test and after Benjamini-Hochberg correction to control the false discovery rate~\cite{benjamini1995controlling}. ***$p < 0.001$, **$p < 0.01$, *$p \le 0.05$}
\begin{tabular}{|p{7.8cm}|p{2cm}|p{2cm}|}
\hline
\textbf{Attribute Level} & \textbf{Mean Acc. Score (\%) of His. Adv.} & \textbf{Mean Acc. Score (\%) of His. Disadv.} \\
\hline
\multicolumn{3}{|c|}{\textbf{Age: Younger (<45) vs. Older (45+)}} \\
\hline
Monetary Incentive: no monetary compensation* & 39.02 & \textbf{46.40} \\
Monetary Incentive: 5\$ gift card* & \textbf{81.20} & 73.74 \\
Monitoring Actor: auditing agencies and regulators* & 72.79 & \textbf{77.22} \\
Monitoring Actor: companies developing automated hiring systems* & \textbf{64.71} & 57.91 \\
\hline

\multicolumn{3}{|c|}{\textbf{Ethnicity: White vs. Non-white}} \\
\hline

%Collected Information Type: Personal data, hiring decisions and qualifications* & 73.85 & \textbf{77.40} \\
Privacy Protection Mechanism: anonymization and encryption*** & 73.61 & \textbf{80.16} \\
Data Storage: specialized commercial companies* & 54.00 & \textbf{60.90} \\
Data Use: develop and evaluate automated hiring systems * & 76.91 & \textbf{80.95} \\
Monitoring Actor: Companies developing automated hiring systems* & 61.49 & \textbf{66.61} \\

\hline

\multicolumn{3}{|c|}{\textbf{Gender: \modtext{Man} vs. Other Genders}} \\
\hline
%Fairness Objective: Select the most qualified candidates* & \textbf{80.67} & 77.30 \\
%Fairness Objective: Treat similar individuals similarly* & \textbf{70.75} & 66.49 \\
Monetary incentive: 5\$ gift card* & \textbf{81.96} & 77.52 \\
\hline

\end{tabular}
\label{tab:acceptance_between_demographic}
\end{table}

We compared protocol design acceptance across demographic groups as shown in Table \ref{tab:acceptance_between_demographic}. 
For different age groups, older participants were more willing to share data without monetary compensation (Mean Acc. Score = 46.4\%) compared to younger participants (Mean Acc. Score = 39.0\%), while younger participants showed higher acceptance of data sharing under \$5 compensation (Mean Acc. Score = 81.2\%) compared to older participants (Mean Acc. Score = 73.7\%). Older participants also expressed greater trust in auditing agencies and regulators as the monitoring actor (Mean Acc. Score = 77.2\%) compared to younger participants (Mean Acc. Score = 72.8\%), while younger participants exhibited higher acceptance of companies developing automated hiring systems in this role (Mean Acc. Score = 64.7\%) compared to older participants (Mean Acc. Score = 57.9\%).

% the mean acceptance score of non-white participants (77.4\%) in contributing all relevant data for fairness monitoring - including personal data, hiring decisions and qualifications - was significantly higher than white participants (73.9\%)

Compared to white participants, non-white participants demonstrated a greater willingness to contribute their data for algorithmic fairness under privacy constraints. In particular, non-white participants (Mean Acc. Score = 80.2\%) were more willing to share their data than white participants (Mean Acc. Score = 73.6\%) when there was only a basic level of privacy protection (i.e., anonymization and encryption). In addition, when commercial companies (as a generally less trusted entity shown in Table \ref{tab:acceptance_dimensions}) took the role of data storage site and monitoring actor, non-white participants demonstrated higher acceptance of data donation (Mean Acc. Score = 60.9\% and 66.6\% respectively) than white-participants (Mean Acc. Score = 54.0\% and 61.5\% respectively). Finally, when data was used to develop and evaluate automated hiring systems, a setting with more direct influence on algorithms yet higher risks of privacy exposure, the mean acceptance score of non-white participants (81.0\%) in contributing their data was significantly greater than white participants (76.9\%).

Regarding gender, \modtext{men} expressed higher acceptance of the protocol with high monetary incentives (Mean Acc. Score = 82.0\%) compared to other participants (Mean Acc. Score = 77.5\%). Besides, other participants valued group fairness (Mean Acc. Score = 75.6\%) more than \modtext{men} (Mean Acc. Score = 72.2\%) with a marginally significant difference (p = 0.052).

% \modtext{man} participants expressed higher acceptance of individual fairness (treat similar individuals similarly, Mean Acc. Score = 70.8\%) and general performance (select the most qualified candidates, Mean Acc. Score = 80.7\%) as the fairness objective compared to non-\modtext{man} participants (Mean Acc. Score = 66.5\% and 77.3\% respectively). In contrast, non-\modtext{man} participants valued more on group fairness (Mean Acc. Score = 75.6\%) compared to \modtext{man} participants (Mean Acc. Score = 72.2\%) with marginally significant difference (p=0.052).

\begin{tcolorbox}[mytakeawaybox]
    We found that users prioritized \textit{collected information type} and \textit{privacy protection mechanisms}, as factors related to privacy risks, in direct attribute rankings. However, in simulated decision-making scenarios, they paid greater attention to benefit-related factors, including \textit{the fairness objective} and \textit{monetary incentive}. According to the protocol design preference, users' privacy concerns manifested in various ways, such as preferring distributed data storage over basic anonymization and encryption, and trusting research centers more than commercial companies to serve as the trusted third party in distributed data storage. Surprisingly, broader data requests and more extensive data use were associated with higher acceptance, possibly due to users' perception of contributing meaningfully to algorithmic fairness. These risk–benefit trade-offs also varied by demographic groups. For instance, non-white participants were significantly more willing than white participants to share data for fairness monitoring under less favorable privacy conditions.
\end{tcolorbox}

\subsection{RQ2: Correlations of Users' Fairness and Privacy Orientations and Contexts with Protocol Design Acceptance}\label{RQ2-findings}

\subsubsection{\modtext{Users open to data sharing prioritize fairness objectives, while privacy-oriented users place less emphasis on fairness objectives or monetary incentives.}}\label{RQ2-benefit}

Table~\ref{tab:factor_benefit} shows how users' fairness and privacy orientations and contexts are associated with the perceived importance of benefit-related attributes based on regression analysis.

For \textbf{fairness objective}, we found that those with greater willingness to donate data ($\beta = 0.24$, $p < .001$) prioritized fairness objective in the ranking task more, indicating that they paid close attention to how their data contributed to social goods. Users with higher levels of privacy protection tended to place slightly less emphasis on the fairness objective associated with data donation. In particular, the perceived importance of data protection transparency had negative correlations with users' attention to fairness objectives during conjoint tasks ($\beta = -0.21$, $p \le .05$), and users with active privacy safeguards tended to rank the fairness objective lower among the attributes of protocol design ($\beta = -0.14$, $p \le .05$).

Regarding \textbf{monetary incentives}, individuals who reported discrimination experienced by close relations were less likely to prioritize compensation benefits when making choices between different protocol designs ($\beta = -0.20$, $p < .01$). Non-binary respondents also expressed lower valuation of monetary incentives in conjoint choices ($\beta = -0.80$, $p \le .05$). Besides, individuals who placed greater importance on data use transparency ($\beta = -0.20$, $p < .001$) and took active privacy safeguards ($\beta = -0.17$, $p < .001$) tended to care less about monetary incentives when ranking the protocol design attributes.

\begin{table*}[htbp]
\small
\centering
\caption{Correlations of Users' Fairness and Privacy Orientations and Contexts with Benefit-Related Attribute. The regression for ``CBC'' is based on individual-level attribute importance derived from Choice-Based Conjoint analysis, which captures revealed attribute importance according to users' choices. The regression for ``Ranking'' is based on the attribute importance derived from users' direct attribute ranking, which reflects stated attribute importance. Inter-Model Reliability denotes the percentage of independent variables whose predictive effects do not conflict between the CBC and Ranking models. ***$p < 0.001$, **$p < 0.01$, *$p \le 0.05$}
\begin{tabular}{|p{6.8cm}|p{1.5cm}p{1.5cm}|p{1.5cm}p{1.5cm}|}
\hline
\textbf{Users' Fairness and Privacy Orientations and Contexts} & \makecell[l]{\textbf{Coef} \\ \textbf{(CBC)}} & \makecell[l]{\textbf{Std. Err.} \\ \textbf{(CBC)}} & \makecell[l]{\textbf{Coef} \\ \textbf{(Ranking)}} & \makecell[l]{\textbf{Std. Err.} \\ \textbf{(Ranking)}} \\
\hline
\multicolumn{5}{|c|}{\textbf{Fairness Objective (Inter-Model Reliability: 85.19\%)}} \\
\hline
Perceived Importance of Data Protection Transparency & -0.21* & 0.10 & - & - \\
General Data Donation Willingness & - & - & 0.24*** & 0.05 \\
Active Privacy Safeguards & - & - & -0.15* & 0.06 \\
Gender: \modtext{Woman} (ref: \modtext{Man}) & -0.07 & 0.07 & -0.09* & 0.05 \\
\hline
\multicolumn{5}{|c|}{\textbf{Monetary Incentive (Inter-Model Reliability: 85.19\%)}} \\
\hline
Perceived Discrimination Experience of Close Relations & -0.20** & 0.07 & -0.10 & 0.05 \\
Gender: Non-binary (ref: \modtext{Man}) & -0.80* & 0.36 & -0.24 & 0.21 \\
Ethnicity: Asian (ref: White) & 0.24* & 0.10 & 0.05 & 0.06 \\
Perceived Importance of Data Use Transparency & -0.17 & 0.09 & -0.20*** & 0.06 \\
Active Privacy Safeguards & -0.10 & 0.09 & -0.17*** & 0.05 \\
Education: Secondary (ref: University) & 0.07 & 0.08 & 0.10* & 0.05 \\
\hline
\end{tabular}
\label{tab:factor_benefit}
\end{table*}

\subsubsection{\modtext{Privacy-oriented users attach more importance to data storage sites and privacy protection mechanisms.}}\label{RQ2-privacy}

Through regression analysis, we also identified how users' fairness and privacy orientations and contexts relate to perceived importance of risk-related attributes, as shown in Table~\ref{tab:factor_privacy}.

Participants who placed greater value on data use transparency assigned more weight to the \textbf{privacy protection mechanism} in their conjoint choices ($\beta = 0.31$, $p < .01$). Likewise, those who emphasized data protection transparency also gave higher importance to this mechanism ($\beta = 0.24$, $p < .001$). In contrast, participants who had experienced discrimination in job applications assigned less importance to the privacy protection mechanism relative to other attributes of the fairness monitoring protocol ($\beta = -0.29$, $p < .01$). 

Finally, in the context of \textbf{data storage}, participants with more active privacy safeguards ($\beta = 0.24$, $p < .01$) and stronger perceived importance of data protection transparency ($\beta = 0.24$, $p \le .05$) placed greater emphasis on where their data would be stored. Compared to white respondents, black respondents showed less concern about data storage ($\beta = -0.33$, $p < .01$).

% Conjoint: (mplicit) Individual-level Attribute Importance derived from users' choices in conjoint tasks. Rank: (stated) attribute importance based on users' attribute ranking. 

\begin{table*}[htbp]
\small
\centering
\caption{Correlations of Users' Fairness and Privacy Orientations and Contexts with Risk-Related Attribute Importance. ***$p < 0.001$, **$p < 0.01$, *$p \le 0.05$}
\begin{tabular}{|p{6.8cm}|p{1.5cm}p{1.5cm}|p{1.5cm}p{1.5cm}|}
\hline
\textbf{Users' Fairness and Privacy Orientations and Contexts} & \makecell[l]{\textbf{Coef} \\ \textbf{(CBC)}} & \makecell[l]{\textbf{Std. Err.} \\ \textbf{(CBC)}} & \makecell[l]{\textbf{Coef} \\ \textbf{(Ranking)}} & \makecell[l]{\textbf{Std. Err.} \\ \textbf{(Ranking)}} \\
\hline
\multicolumn{5}{|c|}{\textbf{Privacy Protection Mechanism (Inter-Model Reliability: 85.19\%)}} \\
\hline
Personal Discrimination Experience in Job Applications & -0.29** & 0.10 & -0.08 & 0.05 \\
Perceived Importance of Data Use Transparency & 0.31** & 0.10 & - & - \\
Perceived Importance of Data Protection Transparency & - & - & 0.24*** & 0.05 \\
Gender: \modtext{Woman} (ref: \modtext{Man}) & 0.11 & 0.07 & 0.10** & 0.04 \\
Ethnicity: Other (ref: White) & -0.09 & 0.12 & -0.15* & 0.06 \\
\hline
\multicolumn{5}{|c|}{\textbf{Data Storage (Inter-Model Reliability: 74.07\%)}} \\
\hline
Ethnicity: Black (ref: White) & -0.33** & 0.11 & -0.04 & 0.06 \\
Active Privacy Safeguards & 0.24** & 0.09 & 0.08 & 0.05 \\
Perceived Importance of Data Protection Transparency & 0.24* & 0.10 & 0.12 & 0.08 \\
\hline
\end{tabular}
\label{tab:factor_privacy}
\end{table*}

\subsubsection{\modtext{Understanding of automated hiring systems increases attention to the collected information type, and marginalized groups give higher priority to ways of data use.}}\label{RQ2-dualImpact}

Table~\ref{tab:factor_dual} presents how users' fairness and privacy orientations and contexts correlate with the importance of design attributes related to both perceived benefits and risks, including \textit{collected information type}, \textit{data use} and \textit{monitoring actor}.

For \textbf{collected information type}, individuals with a stronger understanding of the role of automated hiring systems paid greater attention to this attribute in conjoint decisions ($\beta = 0.27$, $p < .001$). Compared to those with a university-level educational background, individuals with a basic-level education demonstrated significantly greater concerns about which types of personal information were collected in conjoint tasks ($\beta = 0.69$, $p \le .05$). Besides, prior data donation experience positively predicted higher ranks of collected information type among the attributes ($\beta = 0.06$, $p < .01$), whereas individuals with higher willingness to donate data assigned less importance to it ($\beta = -0.15$, $p < .001$).

For \textbf{data use}, respondents identifying as black ($\beta = 0.24$, $p \le .05$), \modtext{woman} ($\beta = 0.17$, $p \le .05$), or non-binary ($\beta = 0.97$, $p < .01$) placed significantly greater importance on how their data would be used. It suggests that historically marginalized groups paid close attention to whether their data had a more direct influence on the algorithmic fairness (evaluate the algorithm vs. evaluate and develop the algorithm). Conversely, individuals who perceived greater privacy risks in data donation were less likely to value data use as an important attribute in conjoint tasks ($\beta = -0.19$, $p \le .05$). Stated importance of data use was negatively associated with older age ($\beta = -0.04$, $p \le .05$), lower educational attainment ($\beta = -0.12$, $p < .01$), and personal discrimination experiences in job applications ($\beta = -0.10$, $p \le .05$). 

% Notably, participants who reported discrimination experienced by close relations ranked data use higher ($\beta = 0.12$, $p < .05$), suggesting that secondhand discrimination may raise awareness about how participants would like to contribute their data for algorithmic fairness.

For the \textbf{monitoring actor}, non-binary individuals ($\beta = 0.74$, $p < .001$) and those with active privacy safeguards ($\beta = 0.23$, $p < .001$) assigned significantly greater importance to this attribute, reflecting their heightened attention to who monitors fairness.

% IAI: (revealed) Individual-level Attribute Importance derived from users' choices in conjoint tasks. Rank: (stated) attribute importance based on users' attribute ranking.

\begin{table*}[htbp]
\small
\centering
\caption{Correlations of Users' Fairness and Privacy Orientations and Contexts with Benefit-Risk Attribute Importance. ***$p < 0.001$, **$p < 0.01$, *$p \le 0.05$}
\begin{tabular}{|p{6.8cm}|p{1.5cm}p{1.5cm}|p{1.5cm}p{1.5cm}|}
\hline
\textbf{Users' Fairness and Privacy Orientations and Contexts} & \makecell[l]{\textbf{Coef} \\ \textbf{(CBC)}} & \makecell[l]{\textbf{Std. Err.} \\ \textbf{(CBC)}} & \makecell[l]{\textbf{Coef} \\ \textbf{(Ranking)}} & \makecell[l]{\textbf{Std. Err.} \\ \textbf{(Ranking)}} \\
\hline
\multicolumn{5}{|c|}{\textbf{Collected Information Type (Inter-Model Reliability: 85.19\%)}} \\
\hline
Understanding of Automated Hiring Systems & 0.27** & 0.08 & -0.07 & 0.05 \\
Education: Basic (ref: University) & 0.69* & 0.32 & 0.17 & 0.18 \\
Prior Data Donation Experience & - & - & 0.06** & 0.02 \\
General Data Donation Willingness & -0.07 & 0.07 & -0.15*** & 0.04 \\
\hline
\multicolumn{5}{|c|}{\textbf{Data Use (Inter-Model Reliability: 70.37\%)}} \\
\hline
Ethnicity: Black (ref: White) & 0.24* & 0.11 & -0.07 & 0.06 \\
Gender: \modtext{Woman} (ref: \modtext{Man}) & 0.17* & 0.07 & 0.07 & 0.04 \\
Gender: Non-binary (ref: \modtext{Man}) & 0.97** & 0.36 & -0.24 & 0.19 \\
Perceived Privacy Risk of Data Donation & -0.19* & 0.09 & 0.07 & 0.05 \\
Age & -0.02 & 0.04 & -0.04* & 0.02 \\
Education: Secondary (ref: University) & -0.08 & 0.08 & -0.12* & 0.05 \\
Personal Discrimination in Job Applications & 0.12 & 0.10 & -0.10* & 0.04 \\
Perceived Discrimination Experience of Close Relations & -0.11 & 0.09 & 0.12* & 0.05 \\
\hline
\multicolumn{5}{|c|}{\textbf{Monitoring Actor (Inter-Model Reliability: 77.78\%)}} \\
\hline
Ethnicity: Other (ref: White) & -0.38** & 0.12 & 0.04 & 0.06 \\
Gender: Non-binary (ref: \modtext{Man}) & - & - & 0.74** & 0.21 \\
Active Privacy Safeguards & - & - & 0.23*** & 0.05 \\
\hline
\end{tabular}
\label{tab:factor_dual}
\end{table*}

% \begin{table*}[ht] \label{ProtocolDesign}
%     \centering
%     \caption{Attributes Presented in the Choice-Based Conjoint Analysis Surveys.}
% \begin{tabular}{|p{2cm}|p{4.5cm}|p{1.8cm}p{1.6cm}|p{1.8cm}p{1.6cm}|}
% \hline
% \textbf{Outcome} & \textbf{Variable} & \textbf{Coef (HB)} & \textbf{Std (HB)} & \textbf{Coef (Rank)} & \textbf{Std (Rank)} \\
% \hline
% \multirow{7}{4cm}{Algorithm\\Influence} 
%  & Nonbinary & 0.596** & 0.195 & -- & -- \\
%  & Ethnicity: Black & 0.121* & 0.055 & -- & -- \\
%  & Willingness to Donate & -- & -- & 0.244*** & 0.054 \\
%  & Education: Secondary & -- & -- & -0.190** & 0.065 \\
%  & Importance: Data Protection & -- & -- & -0.166* & 0.076 \\
%  & Prior Donation Experience & -- & -- & 0.059* & 0.028 \\
%  & Understanding AHS Role & -- & -- & 0.118* & 0.052 \\
% \hline
% \multirow{8}{4cm}{Privacy\\Concerns}
%  & Importance: Data Protection & 0.161** & 0.052 & 0.505*** & 0.087 \\
%  & Importance: Data Use & 0.128** & 0.046 & 0.387*** & 0.078 \\
%  & Discrimination: Close Person & 0.068* & 0.034 & -- & -- \\
%  & Willingness to Donate & -- & -- & -0.326*** & 0.063 \\
%  & Privacy Risk (Donation) & -- & -- & 0.291*** & 0.071 \\
%  & Understanding Info Risks & -- & -- & 0.210* & 0.083 \\
%  & Active Safeguards & -- & -- & 0.135* & 0.068 \\
%  & \modtext{Woman} & -- & -- & 0.130* & 0.065 \\
% \hline
% \end{tabular}

% \vspace{0.5em}
% \footnotesize{\textit{Note:} ***$p < 0.001$, **$p < 0.01$, *$p < 0.05$}

% \end{table*}

\begin{tcolorbox}[mytakeawaybox]
    Regression analysis revealed that users open to data sharing tended to prioritize fairness objectives in the protocol design. In contrast, privacy-oriented users, who exhibit more active privacy behaviors and higher sensitivity to data protection transparency, placed less emphasis on fairness or monetary incentives, focusing instead on risk-related attributes such as data storage and privacy protection mechanisms. Additionally, when choosing a preferred protocol, a better understanding of the algorithmic system positively predicted attention to the collected information type. Respondents identifying as black, \modtext{woman}, or non-binary placed significantly greater importance on how their data would be used for fairness monitoring in conjoint tasks.
\end{tcolorbox}

\section{Discussion}

This study offers empirical insights into how users evaluate and prioritize various design attributes of fairness monitoring privacy-preserving protocols (RQ1), and how these considerations relate to users' fairness and privacy orientations and contexts (RQ2). In this section, we situate the findings within prior work and synthesize implications for the design of future privacy-preserving fairness monitoring protocols.

\subsection{Designing and Communicating Privacy-preserving Fairness Monitoring Protocols: A Human-Centered Perspective}

% This study took a human-centered approach to investigate user preferences for the protocol design of privacy-preserving fairness monitoring across benefit-related, risk-related, and benefit-risk attributes. 

% By examining how users perceive MPC protocol design across multiple design attributes, the findings provide practical implications for \textit{designing} and \textit{communicating} privacy-preserving fairness monitoring protocols from a human-centered perspective.

% Through a human-centered approach to investigate user acceptance of MPC protocol for fairness monitoring

% In particular, participants expressed slightly higher acceptance of the objective of general performance (i.e., selecting the most qualified candidates) compared to \textit{group fairness} and \textit{individual fairness}, although all three goals received over 65\% average acceptance.

 % and \textit{monetary incentive}

%  

\subsubsection{Fairness objectives matter: align fairness goals with outreach strategies to achieve representative datasets}

Fairness objective played a crucial role in shaping users' decisions in conjoint tasks as shown in Section \ref{RQ1-findings}. Notably, \modtext{women} showed stronger support for group fairness than \modtext{men}, aligning with prior findings on differences in accuracy-privacy tradeoffs between historically advantaged and disadvantaged groups~\cite{mourali2025public}. Given the persistent underrepresentation of historically disadvantaged groups in datasets for algorithmic fairness research~\cite{fabris2024fairness}, \textbf{combining targeted outreach with clear fairness objective communications} offers a promising strategy to build more representative datasets. Nonetheless, as algorithmic fairness remains abstract for general users~\cite{woodruff2018qualitative}, it is warranted for future research to investigate effective methods for communicating fairness goals in accessible and engaging ways~\cite{yan2024exploring}. 

\subsubsection{Nuances of monetary incentives: incentivizing participation while avoiding coercion}

Our findings showed that though monetary incentive was ranked as the least important attribute in direct ranking, it played the most important role in influencing users' decision-making in conjoint tasks. Therefore, the effect of monetary incentives in hypothetical scenarios was much greater than in conceptual settings. Together with Berke's work demonstrating even a stronger effect of monetary incentives in real-world settings compared to hypothetical scenarios~\cite{berke2024insights}, we suggest that stakeholders in data donation initiatives treat transparent monetary incentives as an important motivating factor. However, more nuanced considerations are necessary. Our findings indicate that monetary incentives exerted a stronger motivating effect on historically advantaged groups (e.g., \modtext{men} and younger participants) than historically disadvantaged groups, as shown in Section \ref{RQ1-attributeLevels}. Prior work~\cite{ayalon2023exploring,berke2024insights} also highlights that for individuals with intrinsic incentives, monetary incentives may not be the primary participation driver. Therefore, we suggest \textbf{combining moderate monetary incentives with clear communication of fairness objectives}, while \textbf{avoiding excessively high incentives that could risk coercion or distort sample composition with overrepresentation of participants motivated primarily by financial gain}.

\subsubsection{Not less but clearer: focusing on data–impact connections rather than preemptive trade-offs}

Our findings reveal that requesting less data or narrowing data use, often presumed to reduce privacy risks, did not reliably increase acceptance. Specifically, protocols that collected hiring decisions and qualifications in addition to demographic data were more widely accepted than those collecting only demographic data, as shown in Section \ref{RQ1-attributeLevels}. Similarly, participants preferred using data for both development and evaluation over evaluation-only use. These findings suggest that users' privacy calculus hinges on perceived impact and efficacy: when the pathway from specific data items to concrete fairness improvements is made explicit, and paired with credible safeguards and a sense of control, trust and acceptance of the protocol may actually be strengthened rather than diminished. While data minimization remains a critical principle~\cite{ganesh2025data}, we argue that \textbf{when broader data collection or expanded use is necessary, it is more meaningful to foster 
user understanding of how their data enables change through clear 
communication of data-to-impact connections, rather than preemptively 
trading off data collection or usage to promote participation}. To this end, we suggest that future research should actively explore communicating transparent data lifecycles~\cite{polyzotis2018data} to address both data values and risks in support of informed consent.
% function more as a signal of appreciation or a form of effort compensation, rather than serving

% In contrast, historically disadvantaged groups appeared to be more motivated by intrinsic factors, such as contributing to algorithmic fairness. 

\subsubsection{Toward trustworthy MPC protocol design: distributed storage with reliable TTPs}

Our findings reveal nuanced levels of trust toward different stakeholders involved in monitoring or facilitating privacy-preserving data storage. Participants expressed greater acceptance of external monitoring by independent auditing agencies than internal monitoring conducted by companies developing algorithms. Similarly, regarding third parties assisting with distributed data storage, research institutions and NGOs were significantly more trusted than specialized commercial entities. These patterns echo prior research on user trust in neutral or non-stakeholding actors for data donation, such as academic researchers~\cite{berke2024insights}. In terms of privacy protection mechanisms, while baseline approaches like anonymization and encryption already received high acceptance levels, distributed storage further increased acceptance by approximately 10\%, indicating great potential for fairness monitoring under distributed storage in MPC protocols. Given the inherent complexity of coordination and implementation in MPC protocols~\cite{helminger2022multi},  future data donation initiatives should \textbf{prioritize partnering with user-trusted TTPs} like research institutions or NGOs while \textbf{ensuring the technical infrastructure to support secure and scalable MPC implementation}.

\subsection{Connecting Protocol Acceptance with User Contexts and Fairness and Privacy Orientations: Insights for Customizing MPC Protocol Communication}
% communicating nuanced information for informed consent

 % 

% \subsubsection{Demographic factors shape protocol preferences: tailoring protocols for inclusive datasets}

% Demographic characteristics significantly shape individuals' willingness to share data~\cite{berke2024insights,sagvari2021attitudes}. Correlations between demographic characteristics and fairness perceptions~\cite{woodruff2018qualitative,wang2020factors} further complicate this relationship in the context of data donation for fairness monitoring. Our findings enrich the understanding of this relationship. As illustrated in Section \ref{RQ1-attributeLevels}, \modtext{man} and younger respondents were more motivated by monetary incentives, while older participants had higher acceptance of data sharing without monetary compensation. Moreover, older participants showed greater trust disparities between auditing agencies and commercial companies as monitoring actors. Given such demographic differences, prior research has suggested tailoring protocol designs to target demographic needs~\cite{carriere2024best}. For fairness monitoring, we therefore call for protocol designs that accommodate diverse demographic group expectations to ensure inclusive, representative datasets necessary for effective algorithmic fairness measurement~\cite{fabris2022algorithmic}. 

\subsubsection{Engaging historically disadvantaged groups by articulating fairness-related value of contributions}

 % (whether for algorithm evaluation only or also for development), indicating their attenti to sacrifice privacy for algorithmic fairness

We found that demographic characteristics influenced how individuals navigated the tradeoff between contributing to algorithmic fairness and protecting their privacy in data donation. Section \ref{RQ2-dualImpact} reveals that \modtext{women}, black, and non-binary participants paid significantly more attention to data use. Additionally, Section \ref{RQ1-attributeLevels} indicates that non-white participants showed higher acceptance of unfavorable privacy designs for fairness monitoring compared to white participants. It aligns with findings from Section \ref{RQ2-privacy}, showing that respondents who experienced more job application discrimination cared less about privacy protection mechanisms.

 % This includes accepting commercial companies as monitoring entities or data holders, data use for algorithm development beyond evaluation, and weaker privacy protections without distributed data storage as detailed in Section \ref{RQ1-attributeLevels}).

Previous studies have shown that historically disadvantaged groups hold stronger expectations of algorithmic fairness~\cite{mourali2025public,woodruff2018qualitative,wang2020factors}. Extending this literature to data donation settings, our findings demonstrate that individuals from historically disadvantaged backgrounds tend to prioritize fairness goals over privacy concerns, which contributes new insights to inter-group differences in data donation considerations~\cite{berke2024insights,sagvari2021attitudes}. To this end, we call for protocol designs that accommodate diverse demographic group expectations to ensure inclusive, representative datasets necessary for effective algorithmic fairness measurement~\cite{fabris2022algorithmic}. Besides, \textbf{transparently communicating the fairness-related value provides a promising approach to engage historically disadvantaged communities}, such as visual comparisons to show fairness-enhancing effects. However, this approach should focus on objectively informing rather than persuading users to override their privacy concerns. The goal is to enable more informed decision-making by clearly articulating how data contributions advance fairness objectives, while still respecting individual privacy preferences and maintaining genuine informed consent.

\subsubsection{Supporting privacy calculus for meaningful consent: adapting communication to user orientations}

Beyond demographic characteristics, our findings provide insights into how fairness and privacy orientations shape the privacy-benefit calculus. Section \ref{RQ2-benefit} shows that participants with strong privacy orientations, reflected in more privacy behaviors and greater concern for data use and protection transparency, paid significantly less attention to societal or monetary benefits. Instead, they focused more on risk-related attributes like privacy protection mechanisms and collected information types (Section \ref{RQ2-privacy}). Conversely, participants with a higher willingness to share data placed greater emphasis on how their contributions could advance algorithmic fairness. These findings extend the privacy-benefit calculus framework previously examined in contact-tracing app adoption~\cite{geber2021typology,ayalon2023exploring}, which reveals that users with high initial receptiveness focus more on potential benefits, while hesitant users emphasize risks. We recommend that researchers and practitioners design adaptive consent interfaces that respond to users' specific concerns (e.g., AI-powered chatbot ~\cite{xiao2023inform}), \textbf{enabling personalized exploration to meaningfully inform users about donation benefits or available privacy protections while avoiding overwhelming technical complexity}. It is worth noting that customizing consent interfaces based on user orientations requires attention to potential biases from attentional focus. To address this, it is necessary to provide users with comprehensive overviews of critical knowledge sections (such as privacy nutrition labels~\cite{zimmermann2025let}) and highlight information that may be overlooked during their exploration, facilitating access to previously unexamined aspects.

\subsection{Reflecting on the Inconsistency between Stated and Revealed Attribute Importance}

\subsubsection{When \modtext{the specificity of information} shapes informed consent: Toward transparent and concrete protocol communication}

% For instance, as shown in Section \ref{RQ1-attributes}, \textit{monetary incentive} was identified as the most important attribute based on how users compared the design options, yet it was ranked as the least important attribute. More importantly, when users directly engaged in ranking the attributes, they tended to prioritize risk-related attributes (with \textit{collected information type} and \textit{privacy protection mechanism} as the two most important attributes) highly over benefit-related attributes; In contrast, when users were presented the exact protocol designs for comparison, they paid more attention to the benefit side, with \textit{monetary incentive} and \textit{fairness objective} revealed as the top 1 and top 3 important attribute respectively. 

% This work combined stated attribute importance (via direct ranking) with revealed attribute importance (via conjoint choices) to measure how participants assessed and prioritized different design attributes of privacy-preserving fairness monitoring protocols. 

This work reveals the inconsistency between stated and revealed attribute importance when users navigate MPC fairness monitoring protocols. For example, Section \ref{RQ1-attributes} shows that when users ranked attributes without a clear view of how those attributes operate, they tended to prioritize privacy-related features. However, when evaluating complete protocol designs, they shifted their focus toward benefits such as monetary incentives and fairness objectives. This pattern suggests that when users \modtext{were given more specific information about privacy features}, like the exact privacy protection mechanism and collected data type, their attention was more likely to shift to benefits such as fairness contributions. \modtext{These results are in line with the study by Farke et al.~\cite{farke2021privacy}, which observed that increased perceived transparency led users to view the positive sides of data collection}. In fact, rich evidence from other domains suggests that \modtext{providing privacy-related information} does not discourage data donation~\cite{berke2024insights,franzen2024communicating,wang2024role}. \modtext{Nonetheless, mitigating the risk of transparency-washing that pushes for increased unconstrained data collection becomes a crucial topic~\cite{farke2021privacy}.}

This work offers a novel perspective on the privacy-benefit calculus by examining how users evaluate attribute importance differently across varying cognitive levels of risk and benefit understanding. These insights reinforce the importance of \modtext{providing suitable informational depth} for informed consent in data donation initiatives. \textbf{Rather than only communicating ``what the protocol focuses on'', providing more granular, concrete protocol design details, showing ``how the protocol actually addresses these focuses''}, would be a promising way to better enable informed consent by allowing users to make choices that more accurately reflect their values in privacy–benefit calculus.

% It resonated with the study by Farke et al.~\cite{farke2021privacy}, \modtext{which observed that increased perceived transparency through a privacy dashboard led users to view data collection more positively ()}. 

 % through different attribute importance measurements

% Also, users paid more attention to the stakeholder of \textit{data storage} compared to the other two privacy-related attributes (\textit{collected information type} and \textit{privacy protection mechanism}), yet ranked it as the least important among the three privacy-related attributes. 

\subsubsection{\modtext{Draw on both stated and revealed attribute importance} to examine user priorities across design alternatives}

% , which can lead to different judgments

% Methodologically, 
The inconsistency between stated and revealed attribute importance aligns with Dual-Process Theories~\cite{evans2013dual}, which explain human thinking and decision-making as the result of two distinct cognitive systems: System 1 for intuitive processing and System 2 for deliberate processing. \modtext{A possible explanation is that revealed importance (implicitly revealing personal importance of attributes through scenario-based comparisons) aligns more with System 1, while stated importance (systematically comparing abstract attributes to rank their importance) better reflects System 2}. Engaging users with different cognitive processes enables a more comprehensive understanding of how users assess technological protocol designs, particularly in cases of complex decision-making, such as the privacy-benefit calculus examined in this study. We therefore recommend that future researchers and practitioners \modtext{\textbf{draw on both stated and revealed attribute importance to examine user priorities across design alternatives} that could provide complementary insights}.

\subsection{Limitations}

Our study has several limitations. First, we conducted the survey in European countries under GDPR, where legal requirements incentivize the development of data-protection-compliant fairness monitoring protocols~\cite{voigt2017eu,helminger2022multi}. However, both fairness~\cite{berman1985cross} and privacy~\cite{li2022cultural} are sociocultural constructs that vary significantly across societies. We therefore suggest future research investigate user perceptions of privacy-benefit trade-offs to guide privacy-preserving fairness monitoring protocol design across different cultural settings. Second, our study adopted a quantitative approach, employing conjoint analysis and regression models to examine user acceptance of protocol design. Qualitative studies, such as interviews, could further enrich understanding by uncovering how users make sense of the protocol and develop preferences that align with their expectations. Third, while conjoint tasks provide realistic simulation for decision making, they may not fully capture users' perceptions and behaviors in actual data donation contexts~\cite{berke2024insights}. To address this, we recommend continuous user feedback collection during real-world deployments to iteratively refine and adapt privacy-preserving protocols for fairness monitoring. \modtext{Besides, despite our efforts to diversify the participant pool, some marginalized populations, such as non-binary individuals and older job seekers, were still underrepresented in our sample. As these groups are likely to face disadvantages in algorithmic decision-making, ensuring their inclusion in fairness monitoring is vital. More targeted and in-depth research exploring their perspectives on data sharing for privacy-preserving fairness monitoring would therefore be highly valuable. Finally, our work focuses on user perceptions of MPC protocols in the context of fairness monitoring in algorithmic hiring; some findings may not generalize to other data-sharing practices. We suggest future work investigate how users' privacy–benefit trade-offs should inform fairness monitoring protocol designs in other domains (e.g., health, education, migration, etc.).}
% \modtext{Fairness monitoring requires sensitive user data to detect algorithmic bias, which calls for privacy-preserving protocols for user-shared data that align with user expectations.}

 % of data donation under the privacy-preserving protocol

\section{Conclusion}
While data donation offers a promising path for fairness monitoring, developing privacy-preserving protocols that align with users' expectations for data use and privacy protection remains a critical challenge. To address this gap, we conducted a survey study among 833 participants in Europe to examine user acceptance of privacy-preserving protocol design for fairness monitoring. Building on Privacy Calculus Theory and practical multi-party computation protocols for fairness monitoring, we developed design attributes across benefit-related, risk-related, and attributes related to both perceived benefits and risks for investigation. 

First, we assessed design attribute importance through both stated importance via attribute ranking and revealed importance via choice-based conjoint tasks. We found that users prioritized privacy-related features (collected information type and privacy protection mechanisms) in direct evaluations, yet gave greater weight to benefit-related attributes (fairness objective and monetary incentive) when making choices in simulated decision-making scenarios. Users' privacy values manifested in preferred protocol design, including stronger preferences for distributed data storage over basic anonymization and encryption. However, users paradoxically showed greater acceptance of protocols requesting more data and broader data use, potentially due to a stronger sense of contributing to fairness. These benefit-privacy trade-offs varied across demographic groups. For example, non-white participants are more willing to share data for fairness monitoring under less favorable privacy conditions. Besides, regression analysis revealed the correlations between protocol acceptance and users' fairness and privacy orientations and contexts. Users with initial data-donation intentions emphasized fairness objectives more, whereas privacy-oriented users, i.e., those with active privacy behaviors and high sensitivity to data protection transparency, placed less importance on fairness or monetary incentives, focusing instead on privacy features.

Based on these findings, we propose practical implications for privacy-preserving fairness monitoring protocols, covering inclusive protocol design that meets users' expectations and effective protocol communication that enables meaningfully informed consent.

\begin{acks}
This work is supported by the FINDHR project, Horizon Europe grant agreement ID: 101070212. 
\end{acks}

\bibliographystyle{plain}
\bibliography{sample-base}

\appendix

\section{Detailed Explanations of Utility Estimation}\label{Utility-Estimation}

The probability for participant $i$ choosing option $j$ in task $t$ is defined as:

\[
P_{ijt} = \frac{e^{V_{ijt}}}{\sum_{k \in J} e^{V_{ikt}}}
\]

where the total utility $U_{ijt}$ is composed of a systematic part $V_{ijt}$ and an error term:

\[
U_{ijt} = V_{ijt} + \epsilon_{ijt}, \quad \text{with\hspace{0.3cm}} V_{ijt} = \beta_0 + \sum_a \beta_a x_{ajt}.
\]

Here, $\beta_a$ represents the utility weight of attribute $a$, $x_{ajt}$ indicates the level of attribute $a$ for option $j$ in task $t$, and $\epsilon_{ijt}$ captures unobserved variance. We calculated \textit{zero-centered utilities} for each respondent (summing to zero across attribute levels) and \textit{attribute importance scores}, which sum to 100\% per respondent. The relative importance of each attribute was computed as:

\[
I_a = \frac{\max(U_a) - \min(U_a)}{\sum_b (\max(U_b) - \min(U_b))}
\]

This method allowed us to identify how different design attributes influenced users' decision-making regarding the protocol design.

\section{Survey Details}\label{SurveyDetails}
\subsection{BACKGROUND INFORMATION}
    \begin{itemize}
        \item \textbf{What is Data Donation?} 
        \begin{itemize}
            \item Data donation means \textbf{voluntarily sharing personal data} for specific purposes.
            \item It can support \textbf{fairness in Automated Hiring Systems}.
        \end{itemize}
        
        \begin{figure}[htp]
        \centering
        \includegraphics[width=8cm]{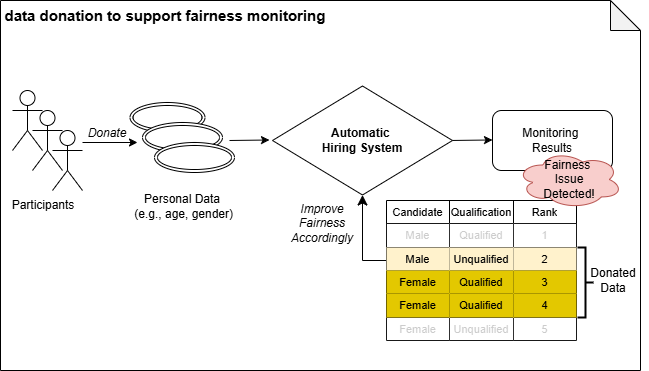}
        \caption{Data Donation to support fairness monitoring}
        \label{fig:DataDonation}
        \end{figure}
        
        \item \textbf{What is an Automated Hiring System?}
        Helps recruiters by:
        \begin{itemize}
            \item Reviewing job applications automatically
            \item Suggesting the best candidates more effectively
        \end{itemize}
        \item \textbf{Why is Fairness Important in Automated Hiring Systems?}
        \begin{itemize}
            \item Automated Hiring Systems can \textbf{inherit biases} from their training data.
            \item Fairness ensures hiring decisions are \textbf{unbiased and non-discriminatory}.
            \item Donated data can help \textbf{reduce biases and improve fairness}.
        \end{itemize}

        \begin{figure}[htp]
        \centering
        \includegraphics[width=8cm]{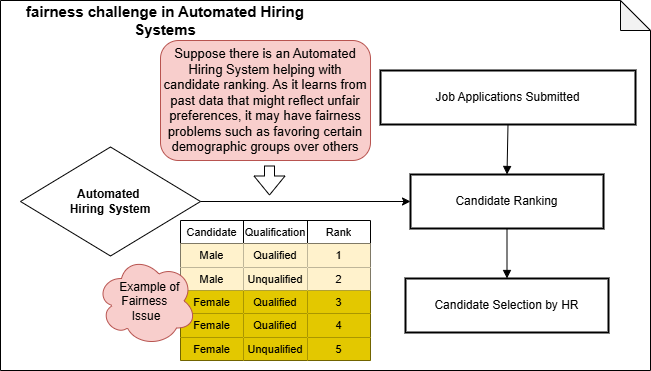}
        \caption{Fairness challenges in Automated Hiring Systems}
        \label{fig:Fairness}
        \end{figure}
        
        \item \textbf{What is Personal Data?}
        \begin{itemize}
            \item Any information that can be used to \textbf{identify you as an individual}
            \item Details about your \textbf{physical or cultural identity}
        \end{itemize}
        \item \textbf{What is Sensitive Data?}
        \begin{itemize}
            \item Sensitive data are a subset of personal data that is \textbf{protected by law more strictly}.
        \end{itemize}
        \item \textbf{Why is protection of Personal Data important in Automated Hiring Systems?}
        \begin{itemize}
            \item Prevents \textbf{misuse of candidate data}.
            \item By \textbf{splitting the data between the third party and the service provider}, neither can access sensitive information independently.
        \end{itemize}

        \begin{figure}[htp]
        \centering
        \includegraphics[width=8cm]{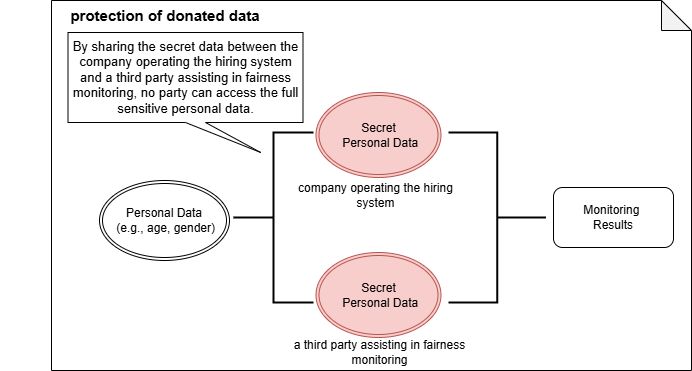}
        \caption{Protection of donated data}
        \label{fig:DataProtection}
        \end{figure}
        
    \end{itemize}

\subsection{Evaluation Questions}
After having read the overview, please answer the following questions.\footnote{Correct Answer: No, Yes, Yes}
\begin{itemize}
    \item Automated Hiring Systems always make fair decisions. (Yes/No)
    % \begin{itemize}
    %     \item Yes
    %     \item No
    % \end{itemize}
    \item Data Donations can support the monitoring of fairness problems (Yes/No)
    % \begin{itemize}
    %     \item Yes
    %     \item No
    % \end{itemize}
    \item There exists a fairness monitoring solution where a third party can monitor fairness without anyone storing the full personal data of job candidates. (Yes/No)
    % \begin{itemize}
    %     \item Yes
    %     \item No
    % \end{itemize}
\end{itemize}

\subsection{DEMOGRAPHIC QUESTIONS}
\begin{itemize}
    \item What is your age? [Under 18/18-24/25-34/35-44/45-54/55-64/65+]

    \item What is your gender? [Male/Female/Non-binary/Prefer not to say/Other]
    
    \item What is your ethnicity (Select all that apply) [White (e.g., European, Middle Eastern, North African)/Black (e.g., African, Afro-Caribbean)/Asian (e.g., South Asian, East Asian, Southeast Asian)/Mixed or Multiple ethnic groups/Prefer not to say/Other]

    \item What is your highest education level? [No formal education/Primary education/Lower secondary education (e.g., GCSE, Junior Certificate)/Upper secondary education (e.g., A-levels, Baccalaureate, Leaving Certificate)/Vocational or technical training/Bachelor's degree (or equivalent)/Master's degree (or equivalent)/Doctorate (PhD or equivalent)/Other]

    \item What is your employment status?[Employed/Unemployed/Prefer not to say]

    \item How many times have you participated in data donations for research or social good? [Never/Once/2-3 times/4-5 times/More than 5 times]
\end{itemize}

\subsection{LIKERT SCALE QUESTIONS}
\begin{itemize}

\item To what extent have you personally experienced discrimination in any context? [-3 (Not at all) - 3 (To a great extent)]
% [-3 (Not at all), -2, -1, 0, 1, 2, 3 (To a great extent)]

\item To what extent do you think you personally experienced discrimination in the process of applying for jobs? [-3 (Not at all) to 3 (To a great extent)]
% [-3 (Not at all), -2, -1, 0, 1, 2, 3 (To a great extent)]

\item To what extent have people close to you (e.g., family, friends, relatives) experienced discrimination when applying for jobs? [-3 (Not at all) to 3 (To a great extent)]
% [-3 (Not at all), -2, -1, 0, 1, 2, 3 (To a great extent)]

\item How knowledgeable are you about the potential risks and consequences of sharing your personal information online? [-3 (Not knowledgeable at all) to 3 (Extremely knowledgeable)]
% [-3 (Not knowledgeable at all), -2, -1, 0, 1, 2, 3 (Extremely knowledgeable)]

\item How knowledgeable are you about how your personal data is collected, used, shared, and protected by the digital services you use daily? [-3 (Not knowledgeable at all) to 3 (Extremely knowledgeable)]
% [-3 (Not knowledgeable at all), -2, -1, 0, 1, 2, 3 (Extremely knowledgeable)]

\item How often do you take active measures to safeguard your personal information online? [-3 (Not at all) to 3 (To a great extent)]
% [-3 (Not at all), -2, -1, 0, 1, 2, 3 (Always)]

\item How knowledgeable are you about fairness monitoring systems designed to detect discrimination in decision-making processes? [-3 (Not knowledgeable at all) to 3 (Extremely knowledgeable)]
% [-3 (Not knowledgeable at all), -2, -1, 0, 1, 2, 3 (Extremely knowledgeable)]

\item This is an attention-check question. Please identify and select the number three (3) from the options below. [-3 to 3]
% [-3 (Greatly negative), -2, -1, 0, 1, 2, 3 (Greatly positive)]

\item How knowledgeable are you about the role of Automated Hiring Systems in hiring processes, for example in processing job applications? [-3 (Not knowledgeable at all) to 3 (Extremely knowledgeable)]
% [-3 (Not knowledgeable at all), -2, -1, 0, 1, 2, 3 (Extremely knowledgeable)]

\item How important is it for you to understand how exactly your personal data will be used if you donate it? [-3 (Not important at all) to 3 (Extremely important)]
% [-3 (Not important at all), -2, -1, 0, 1, 2, 3 (Extremely important)]

\item How important is it for you to understand how your personal data will be protected if you donate it? [-3 (Not important at all) to 3 (Extremely important)]
% [-3 (Not important at all), -2, -1, 0, 1, 2, 3 (Extremely important)]

\item How fair do you think current Automated Hiring Systems are when processing job applications during recruitment? [-3 (Not fair at all) to 3 (Extremely fair)]
% [-3 (Not fair at all), -2, -1, 0, 1, 2, 3 (Extremely fair)]

\item How willing are you to donate your personal data for research or other purposes? [-3 (Not at all) to 3 (Very willing)]
% [-3 (Not at all), -2, -1, 0, 1, 2, 3 (Very willing)]

\item How effective do you think current fairness monitoring systems are at detecting and reducing bias and discrimination in decision-making processes? [-3 (Not effective at all) to 3 (Very effective)]
% [-3 (Not effective at all), -2, -1, 0, 1, 2, 3 (Very effective)]

\item How risky do you think donating your data would be to your privacy? [-3 (Not risky at all) to 3 (Extremely risky)]
% [-3 (Not risky at all), -2, -1, 0, 1, 2, 3 (Extremely risky)]

\end{itemize}

% \newpage
\subsection{CONJOINT CHOICE TASK}

Five conjoint tasks to indicate their preference for protocol design, with an example shown in Figure \ref{fig:conjoint}.
% \begin{figure*}[htbp]
%     \centering
%     \includegraphics[width=0.9\textwidth]
%     {MPC_pics/conjointExample.png}
%     \caption{Conjoint choice task between two fairness monitoring systems}
%     \label{fig:conjointExample}
% \end{figure*}

\subsection{RANKING QUESTION}

Drag to rank the seven attributes — fairness objective, monetary incentive, collected information type, privacy protection mechanism, data storage, monitoring actor, and data use.
% \begin{figure*}[htbp]
%     \centering
%     \includegraphics[width=0.9\textwidth]{MPC_pics/rankingQuestion.png}
%     \caption{Ranking dimensions by user-perceived importance}
%     \label{fig:rankingQuestion}
% \end{figure*}

\subsection{GRID QUESTIONS OF LEVEL OF ACCEPTANCE OF EACH OPTION OF STUDIES DIMENSIONS}

In the next 7 questions, you will rate your level of acceptance for each option within the studied dimensions.
A rating of \textbf{-3} indicates the least preferred option, while \textbf{+3} indicates the most preferred.

\begin{itemize}

\item \textbf{How data donation helps improve Automated Hiring Systems?}:
\begin{itemize}
    \item Select the most qualified candidates
[-3 to 3]
    \item Treat different demographic groups similarly
    [-3 to 3]
    \item Treat similar individuals similarly
    [-3 to 3]
\end{itemize}

\item \textbf{Who is collecting data?}: 
\begin{itemize}
    
    \item Auditing agencies and regulators
    [-3 to 3]
    \item Companies developing automated hiring systems
    [-3 to 3]
\end{itemize}

\item \textbf{How is donated data used?}: 
\begin{itemize}
    \item Evaluate the Automated Hiring System
    [-3 to 3]
    \item Develop and evaluate the Automated Hiring System
    [-3 to 3]
\end{itemize}

\item \textbf{How is data donor compensated?}: 
\begin{itemize}
    \item No monetary compensation
    [-3 to 3]
    \item 1\$ gift card
    [-3 to 3]
    \item 5\$ gift card
    [-3 to 3]
\end{itemize}

\item \textbf{Which data is collected?}: 
\begin{itemize}
    \item Personal data
    [-3 to 3]
    \item Personal data and hiring decisions
    [-3 to 3]
    \item Personal data, hiring decisions and qualifications
    [-3 to 3]
\end{itemize}

\item \textbf{How is data protected?}
\begin{itemize}
    \item Anonymization and Encryption
    [-3 to 3]
    \item Anonymization, Encryption, and Distributed Storage
    [-3 to 3]
\end{itemize}

\item \textbf{who helps to store the encrypted data in separate parts?}
\begin{itemize}
    \item Specialized commercial companies
    [-3 to 3]
    \item Research centers
    [-3 to 3]
    \item Non-governmental non-profit organizations
    [-3 to 3]
\end{itemize}

\end{itemize}

\subsection{OPEN-END QUESTION}
Any feedback or comments on the survey content, design, or anything related you'd like to share?

% \section{Comparison of different demographic groups regarding CBC-based and ranking-based Attribute Importance}
% \begin{figure*}[htbp]
%     \centering
%     \includegraphics[width=0.95\textwidth, height=8cm]{MPC_pics/Demographic_Comparison_Plots/Comparison_of_Attribute_Importance_in_MPC_monitoring_protocol_of_different_Age_groups_plot.pdf}
%     \caption{Comparison of Attribute Importance in MPC monitoring protocol of different Age groups according to Choice-based conjoint and Ranking Tasks}
%     \label{fig:compare_cbc_age}
% \end{figure*}

% \begin{figure*}[htbp]
%     \centering
%     \includegraphics[width=0.95\textwidth, height=8cm]{MPC_pics/Demographic_Comparison_Plots/Comparison_of_Attribute_Importance_in_MPC_monitoring_protocol_of_different_Ethnicity_groups_plot.pdf}
%     \caption{Comparison of Attribute Importance in MPC monitoring protocol of different Ethnicity groups according to Choice-based Conjoint and Ranking tasks}
%     \label{fig:compare_rank_age}
% \end{figure*}

% \begin{figure*}[htbp]
%     \centering
%     \includegraphics[width=0.95\textwidth, height=8cm]{MPC_pics/Demographic_Comparison_Plots/Comparison_of_Attribute_Importance_in_MPC_monitoring_protocol_of_different_Gender_groups_plot.pdf}
%     \caption{Comparison of Attribute Importance in MPC monitoring protocol of different Gender groups according to Choice-based conjoint and Ranking tasks}
%     \label{fig:compare_cbc_ethnicity}
% \end{figure*}

\end{document}